\documentclass{article}

\usepackage{graphicx}
\usepackage{subfigure}
\usepackage{amsmath}
\usepackage{dsfont}
\usepackage{multirow}
\usepackage{fancyhdr}
\usepackage{wrapfig}

\title{Improving the Security of Image Manipulation Detection through
	One-and-a-half-class Multiple Classification}
\author{Mauro Barni, Ehsan Nowroozi and Benedetta Tondi \\
	University of Siena, Italy \\ ehsan.nowroozi@diism.unisi.it \\ ehsan.nowroozi65@gmail.com \footnote{The 1.5C code is avaliable on: http://github.com/EhsanNowroozi/    Accessed: 2019-09-05. Corresponding author: Ehsan Nowroozi}}

\begin{document}
\maketitle
\begin{abstract}
	Protecting image manipulation detectors against perfect knowledge attacks requires the adoption of detector architectures which are intrinsically difficult to attack. In this paper, we do so, by exploiting a recently proposed multiple-classifier architecture combining the improved security of 1-Class (1C) classification and the good performance ensured by conventional 2-Class (2C) classification in the absence of attacks. The architecture, also known as 1.5-Class (1.5C) classifier, consists of one 2C classifier and two 1C classifiers run in parallel followed by a final 1C classifier. In our system, the first three classifiers are implemented by means of Support Vector Machines (SVM) fed with SPAM features. The outputs of such classifiers are then processed by a final 1C SVM in charge of making the final decision. Particular care is taken to design a proper strategy to train the SVMs the 1.5C classifier relies on. This is a crucial task, due to the difficulty of training the two 1C classifiers at the front end of the system. We assessed the performance of the proposed solution with regard to three manipulation detection tasks, namely image resizing, contrast enhancement and median filtering. As a result the security improvement allowed by the 1.5C architecture with respect to a conventional 2C solution is confirmed, with a performance loss in the absence of attacks that  remains at a negligible level.
\end{abstract}

\section{Introduction}

The development of secure image forensics tools, capable of granting good performance even in the presence of an adversary aiming at impeding a correct analysis is receiving increasing attention, due to the ease with which Counter-Forensics (CF) tools erasing the weak traces the image forensic analysis relies on can be designed \cite{Gloe07}. This is even more the case with image forensics techniques based on Machine Learning (ML). In such a case, in fact, the adversary can exploit the weakness of machine learning tools against intentional attacks as discussed in \cite{huang2011adversarial} and shown in \cite{barni2018adversarial} for the particular case of ML-based image forensic tools.

The kind of techniques the adversary can use to launch his attacks, and consequently the countermeasures the forensic analyst can rely on to combat them, depends heavily on the information that the adversary has about the to-be-attacked system \cite{barni2018adversarial}.  By following \cite{Biggio2013evasionAttack}, we say that we are in a Perfect Knowledge (PK) scenario, when the adversary has a complete information about the forensic algorithm under attack. In the case of techniques based on ML, this includes the exact knowledge of the ML model resulting from the training process.

In the PK case, very powerful CF attacks can be launched, which are capable of preventing a correct analysis while introducing a limited distortion into the attacked image. Designing effective anti-CF solutions in a PK setting is a difficult task, since in this scenario it is the attacker who plays the last move. The PK assumption, in fact, implies that the attacker is aware of any possible countermeasure adopted by the forensic analyst. The only solution in this case is to rely on image forensics techniques which are intrinsically more difficult to attack.

In this paper, we focus on image manipulation detection, a situation in which the forensic analysis aims at deciding whether a given image has been subject to a certain manipulation or not. In this case, the image forensic analysis boils down to a binary decision, according to which the image space is partitioned into two regions, corresponding, respectively,  to pristine and manipulated images. The most classical approach to face with this problem in a ML setting, is to train the manipulation detector by presenting to it a number of images belonging to the two classes. The goal of the training process is to partition the image space in such a way that the training examples are classified correctly. The decision rule adopted by the detector must be designed in such a way to maintain good performance also when the detector is faced with new images that were not contained in the training set (that is the detector must have good generalization capabilities).

A problem with the approach described above is that the detector is forced to make a decision for one of the two classes of images even when it is faced with images whose characteristics  are far away from those of the images seen during training, possibly because such images do not genuinely belong to any of the two classes. Suppose, for instance, that a classifier has been trained to distinguish between pristine images directly acquired by a photocamera with no further processing and median filtered images \cite{yuan2011blind}. As long as the images fed to the classifier belong to one of the above classes, the classifier will provide a correct answer, based on the examples it has {\em seen} during training. However, if we ask the classifier to make a decision on an image with completely different characteristics, e.g. a line drawing, a cartoon, or a bilevel image, what would the classifier do? Very likely it would make a kind of random decision according to the way the partition of the image space optimised on the images of the training set is extended to regions of the image space for which no training example was provided.
In normal working conditions this may not be a problem, since one can assume that the classifier will never be asked to work in such abnormal conditions. In an adversarial setting, however, the presence of unpopulated regions of the image space can be exploited by the attacker to induce a classification error at a low price (image distortion). The above situation is depicted in Fig. \ref{fig220:setup}(a), where a point of the red class is moved into an unpopulated part of the blue-decision region with a minimum effort.

When the goal of the adversary is a unidirectional one, that is, when the adversary aims at attacking only images belonging to one of the two classes, it is possible to cope with the problem outlined above by resorting to a 1-Class (1C) classifier \cite{Tax01}. In this case, the ML model is trained with images belonging to one class only, so to teach it to recognise if an image belongs to such a class or not. In this scenario, all other kinds of images will be classified as extraneous ones. The main difference between a conventional 2-Class (2C) classifier and a 1-class one, is that, in the latter case, the image space is partitioned into a closed region, containing the images belonging to the class used during training, and a complementary region with all the other images. As shown in Fig. \ref{fig220:setup}(b), an attacker aiming at moving an image in the outside (red) region into the closed region characterising the images of the target class (blue region) can no more exploit the presence of unpopulated regions of the image space since such regions are - by default - assigned to the outer space.

Despite its attractiveness from a security point of view, the use of 1C classification has its own drawbacks, deriving from the fact that 1C classifiers do not exploit  any knowledge about the images  belonging to one of the two classes. By considering again the example of pristine and median filtered images, a 1C classifier could be trained by considering only pristine images, however, in this way it would not exploit any information about the particular traces left within an image by a median filter. As a result, the performance of 1C classifiers in the absence of attacks are expected to be lower than those obtainable by means of a 2C architecture.\\

In order to couple the superior accuracy achievable by 2C classifiers and the intrinsic security of 1C classifiers, Biggio et al. \cite{biggio2015one} has proposed a multiple classifier architecture which combines the advantages of both approaches. The proposed classifier, called one-and-a-half Class (1.5C) classifier, is therein applied to spam and malware detection.

In this paper, we propose a 1.5C architecture for image manipulation detection and assess its effectiveness in the presence of ad-hoc CF attacks in a PK scenario. More specifically, we consider three different kinds of manipulations: image resizing, median filtering, and contrast enhancement, and show how the 1.5C classifier can be designed in such a way to achieve good performance in the absence of attacks (comparable to those achieved by a standard 2-C classifier) while at the same time being more difficult to attack in a PK scenario, thus achieving a better security with respect to the 2C case. Particular care is paid to the training process, since training the 1.5C classifier turns out to be a difficult task due to the difficulty of training the 1C classifiers it consists of. The security improvement allowed by the 1.5C architecture with respect to a conventional 2C solution is confirmed experimentally, with a performance loss in the absence of attacks that  remains at a negligible level.

The rest of the paper is organised as follows.
In Section \ref{sec.part}, we describe previous attempts to improve the security of image forensics analysis by resorting to 1C classification. Then, in Section \ref{prop_system}, we introduce the proposed 1.5C classifier for image manipulation detection and describe the strategy we followed to train it. In Section \ref{sec.methodology}, we describe the methodology we used to evaluate the effectiveness of the proposed system, while in Section \ref{sec.results}, we present the corresponding experimental results. The paper ends in Section \ref{sec.con} where we summarise the main results of our research and suggest some directions for future works.

\label{sec.results}

\section{Prior art}
\label{sec.part}

The use of 1C classifiers is not novel in multimedia forensics and security-related applications.
In \cite{d2017autoencoder}, the authors resort to 1C classification for video forgery detection,
in the attempt to devise a tool which works under challenging  conditions, like those encountered in social networks.
In particular, the authors resort to the use of an architecture based on {\em autoencoders} trained on pristine data. When used in this way, autoencoders behave as 1C classifiers, with a large reconstruction error between the input and output being interpreted as an anomaly, i.e. a forgery.  More in general, one class modelling is popular for anomaly detection in many different applications, where a good statistical characterisation under abnormal condition is not available. As an example, we mention the problem of acoustic novelty detection \cite{acusticDet} or the problem of detection of abnormal events in video sequences \cite{hasan2016learning}.  A method for adversarial anomaly detection, which exploits the combination of multiple 1C classifiers to increase the hardness of evasion attacks against intrusion detection systems is provided in \cite{EnsembleOneClassIDS}.

1C classification has often been proposed as an alternative solution to conventional multi-class algorithms, that classify a sample based on a number of pre-defined categories, when an exhaustive list of such categories does not exist \cite{SurveyOneClass}. This problem is often referred to as classification in {\em open set} conditions. Open set problems have been studied in several image forensic and security-oriented applications,  such as fingerprint spoof detection \cite{OpenSetSpoof} source device attribution \cite{OpenSetScheirer}, and camera model identification \cite{Wang2009,inproceedings}. In \cite{Wang2009}, in particular, a combination of one-class and multi-class SVMs is used to simultaneously recognize the camera model among the models in a known set  and, at the same time, identify outliers, acquired by unknown camera models.

More recently, 1C classifiers have been used in conjunction with generative adversarial models (GANs) to design detectors with work under the assumption that no or very few instances of malicious samples are available for training. This is the case in \cite{yarlagadda2018satellite}, where  the problem of forgery detection of satellite imagery is addressed, and in \cite{zheng2018one}, with regard to general fraud detection, e.g. in reputation systems.

\section{Proposed System}
\label{prop_system}
In this section, we first formalise the detection problem addressed in this paper, then we describe the 1.5C architecture originally proposed in \cite{biggio2015one} for pattern recognition applications, the choice of the feature set, and the methodology that we followed to train the 1.5C classifier.

\begin{figure}[t!]
	\centering
	\subfigure[]{\includegraphics[width=0.48\textwidth , height = 2.5cm]{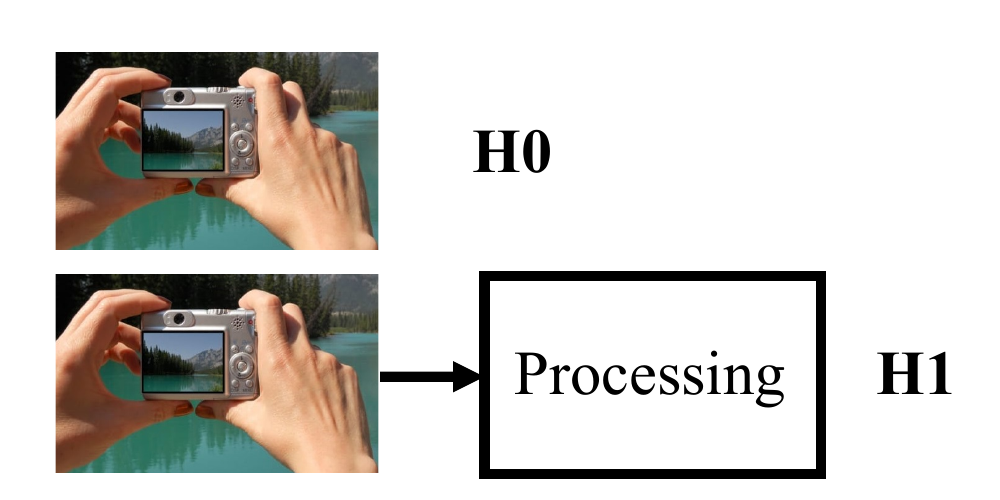}}\\
	\subfigure[]{\includegraphics[width=0.7\textwidth , height = 3.5cm]{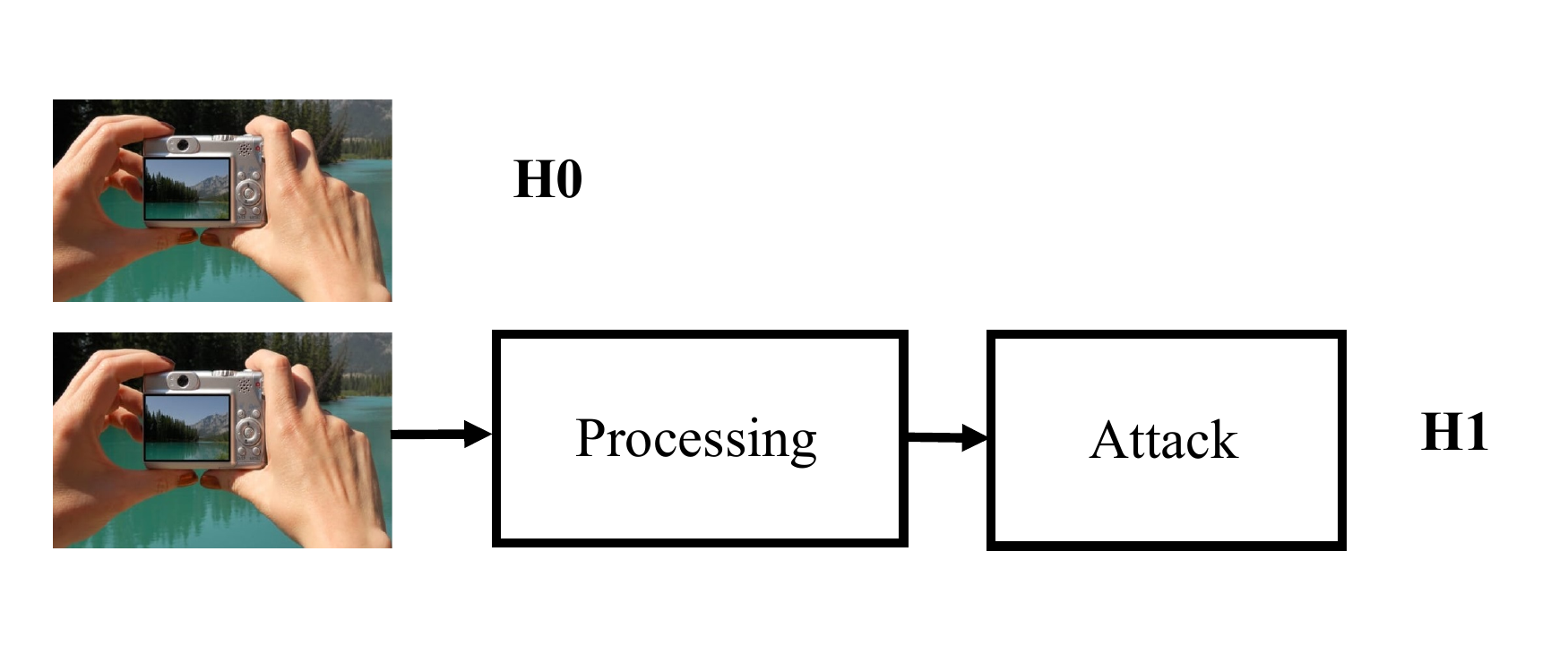}}
	\caption{General manipulation detection task considered in this paper (a) and its adversarial version (b).}
	\label{Fig1.setup}
\end{figure}

The basic detection task addressed in this paper is schematised in Fig.  \ref{Fig1.setup}(a).

Hypothesis $H_0$ corresponds to the case of pristine images produced by an acquisition device without any subsequent processing. $H_1$ corresponds to the case of processed or manipulated images. Fig. \ref{Fig1.setup}(b) depicts the case in which the same detection task is carried out in an adversarial setting, in the presence of an attacker aiming at impeding a correct detection. Specifically, we assume that the goal of the attacker is to avoid that the manipulation is detected, that is, the attacker's goal is to induce a missed detection error. This kind of attack, usually referred to as an {\em integrity violation} attack \cite{huang2011adversarial}, is the most common one in counter-forensics. In the sequel, we denote with $P_{md}$ the probability of a missed detection error, namely the probability that a manipulated image is detected as a pristine image (also indicated as $P(H_0|H_1)$), and with $P_{fa}$ the false alarm probability, that is the probability that a pristine image is detected as being manipulated (namely, $P(H_1|H_0)$).
\subsection{Architecture of the One-and-a-half-class (1.5C) Classifier }
\begin{figure}[t!]
	\centering
	\includegraphics[width=1.1\textwidth , height = 6.5cm]{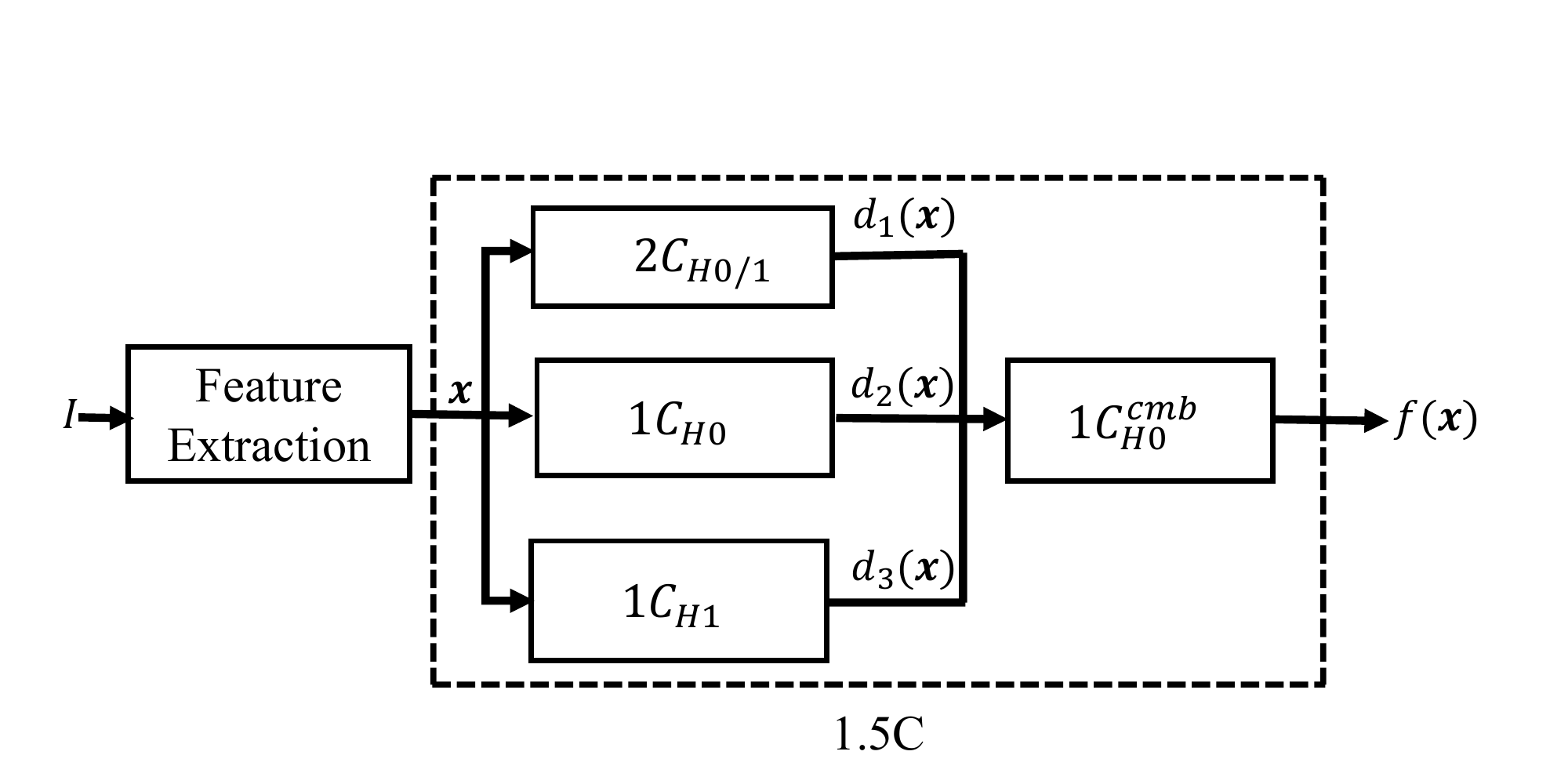}
	\caption{Architecture of the one-and-a-half class (1.5C) classifier adopted in this paper. $I$ is the input image, while ${\bf x}$ is the feature vector extracted from $I$.}
	\label{Fig2.setup}
\end{figure}

The architecture of the one-and-a-half-class classifier adopted in this paper is depicted in Fig. \ref{Fig2.setup}.
Given an image $I$, a feature vector ${\bf x}(I)$, or simply ${\bf x}$, is extracted and used to feed a multiple classifier whose first stage consists of the parallel application of three classifiers: a 2C classifier, trained with examples belonging to both classes (in our case, pristine and manipulated images) and two 1C classifiers, one trained with images belonging to the $H_0$ class (pristine images) and the other trained with images from the $H_1$ class (manipulated images). The outputs of these classifiers represent the input of a final 1C classifier, referred to as combination classifier. Since the goal of our work is to increase the security in the presence of integrity violation attacks, the final classifier is trained with pristine data.

We denote with $d_1({\bf x}),d_2({\bf x})$ and $d_3({\bf x})$ the output respectively of the 2C, 1C trained on pristine images and
1C  trained on manipulated images classifiers;  $f({\bf x})$ denotes the decision function of the downstream 1C combination classifier trained with pristine images. For ease of reference, in the sequel we indicate the above classifiers with the acronyms $2C_{H_{0/1}}, 1C_{H_0}, 1C_{H_1}$ and $1C_{H_0}^{cmb}$, as indicated in Fig. \ref{Fig2.setup}.

\begin{figure}
	\centering
	
	\subfigure[2-C]{\label{fig220:a}\includegraphics[width=0.32\columnwidth]{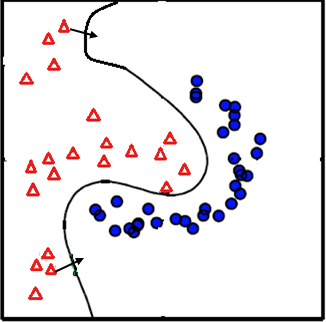}}
	\subfigure[1-C]{\label{fig220:b}\includegraphics[width=0.32\columnwidth]{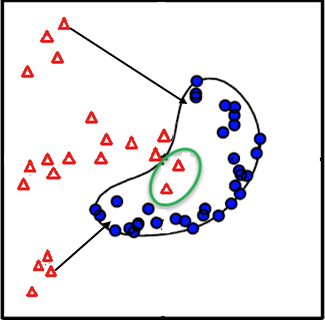}}
	\subfigure[1.5C]{\label{fig220:c}\includegraphics[width=0.32\columnwidth]{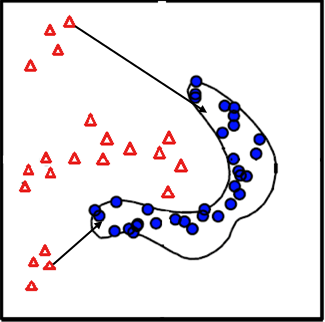}}
	\caption{Pictorial representation of the decision regions for 2C, 1C and 1.5C classifiers. Blue dots refer to $H_0$ hypothesis, red triangles to $H_1$. The goal of the attacker is to take samples belonging to $H_1$ detection region and move them inside the region for which the detector decides in favour of $H_0$. The performance of the 1.5C classifier are in line with those of the 2C, while it is expected to be more robust under attacks, since a larger distortion, exemplified in the picture by a longer arrow, is needed to move a sample from the $H_1$ to the $H_0$ decision region. The presence of red triangles within the $H_0$ decision region in part b) illustrates the lower performance of the 1C classifier in the absence of attacks.}
	\label{fig220:setup}
\end{figure}

As discussed in the introduction, 2C classifiers are often capable of achieving high accuracy in the absence of attacks, however they do not generalize well to examples that were not properly represented in the training phase, thus easing the task of the adversary, that can exploit the presence of unexplored regions of the features space to carry out his attack. Such a situation is exemplified in part (a) of Fig. \ref{fig220:setup}, where the attacker exploits the presence of empty regions to induce a missed detection error (for simplicity, the figure refers to a case of perfect classification in the absence of attacks). As a result, the 2C classifier  may be easily attacked as shown for instance in \cite{CFharder16,marra2015counter} in the case of image manipulation and forgery detection. On the other hand, by defining a closed region enclosing the samples from one class only - usually the $H_0$ class, 1C classifiers are intrinsically more robust against attacks, even if they may get worse performance in the absence of attacks. This behavior is illustrated in Fig. \ref{fig220:setup} (b), where we see that moving a sample from the $H_1$ to the $H_0$ region requires a larger distortion, due to the closeness of the acceptance region for $H_0$. Such an advantage comes at the price of a reduced accuracy in the absence of attacks, since the one class classifier is not able to properly shape the closed acceptance region given that it is trained only on examples generated under $H_0$ (the samples for which a missed detection error occurs are highlighted by a circle in the figure).

The goal of the 1.5C classifier is to simultaneously retain the advantages of 2C and 1C classification, as illustrated in  Fig. \ref{fig220:setup} (c). In particular, the 1.5C classifier is expected to have a similar robustness  against attacks as the 1C classifier (moving a sample from  $H_1$ to $H_0$ requires a similar distortion), while the acceptance region is better shaped with respect to the 1C classification case, then the performance in the absence of attacks are improved compared to the 1C classifier, and are more similar to those obtained by the 2C classifier  (the same perfect classification is achieved in the illustrative example in the figure). This behaviour is confirmed in our experiments for all the manipulation detection tasks we have considered (see the results in Section \ref{sec.performance_No_A} and \ref{sec.performance_A}).
\subsection{Implementation of 1.5C Detector and Choice of the Feature Set}
\label{sec.featureSelection}

The 1.5C architecture outlined in the previous section is a general one, and, in principle, can be implemented by resorting to any kind of 1C and 2C classifiers. In this work, we decided to implement all the classifiers the 1.5C classifier consists of by means of Support Vector Machines (SVMs). The main reason is the ease with which SVMs can be used to implement 1C classification. It goes without saying that the use of alternative architectures can be explored as a future research direction.

With regard to the choice of the feature set, we tried to balance between two opposite requirements. On one hand, the features have to be powerful enough to capture the different types of dependencies among neighbouring pixels in pristine and manipulated images. On the other hand, we want to keep the dimensionality of the feature set limited, so to make it possible to design the intermediate and combination classifiers as SVMs, without resorting to more complicated architectures such as ensemble classifiers \cite{kodovsky2012ensemble}.
Specifically, we selected the Substractive Pixel Adjancency Model (SPAM) feature set  \cite{SPAM686}, originally proposed for steganalysis and extensively used in image forensic applications \cite{verdoliva2014feature,li2018identification}. The SPAM features are extracted as follows: first, the first-order residuals in all the directions (horizontal, vertical, diagonal) and orientations (left to right and right to left, up to down and down to up) are computed $\{  \leftarrow , \to , \uparrow , \downarrow , \nwarrow , \searrow , \swarrow , \nearrow \}$; for instance, for the horizontal left to right direction we have $D_{i,j}^ \to  = {I_{i,j}} - {I_{i,j + 1}}$. Then, the residual values are truncated at $T$ (by default, $T=3$); finally, the second-order co-occurrences of the truncated residuals are computed.
The number of features is reduced by exploiting the symmetries so that the dimension of the final feature vector is equal to 686.  We refer to  \cite{SPAM686} for the details of the feature computation procedure. SPAM features are designed for grayscale images; when working with color images, they can be extracted from the luminance channel.

\subsection{Training the 1.5C Classifier}
In this section, we describe the strategy we have adopted to train the intermediate and the combination classifiers. In general, training the 1.5C classifier is not  easy, especially when the detection task is not straightforward, as it is often the case in image forensic applications. The difficulties are mainly associated to the one-class classifiers which are difficult to train and may not achieve good classification performance in many cases. The four classifiers of the 1.5C system are all SVMs whose hyper-parameters are determined by means of a preliminary internal validation phase. Let $({x_i}, y_i)$ be the training pair for image $i$, where ${x_i} \in \mathds{R}^r$ denotes the feature vector of dimensionality $r$ and $y_i$ denotes the class label associated to the image.

For a general intermediate SVM (hereafter indicated by the index $j \in \{2C_{H_{0/1}}, 1C_{H_0}, 1C_{H_1}$\}), the decision function $d_j(x)$ learned by the classifier is expressed as:
\begin{equation}
\label{dec_function}
d_j(x) = \sum\limits_{i = 1}^n {{y_i}{\alpha_{j,i}}K_j({x_i},x) + b_j }
\end{equation}
where $n$ is the number of training images, $K_j({x_i},x)$ is the kernel function of the SVM, $b_j$ is the bias term, and ${\bf \alpha}_j$ is a vector of scalars with $0 \le \alpha_{j,i} \le C_j$  where $C_j$ is the cost term, that is, the penalty parameter of the error term (over the training set) of the SVM optimization problem \cite{boser1992training}. In our case, $x_i$ is the SPAM feature vector of the $i$-th image; then, $r = 686$ and $x_i \in \mathds{R}^{686}$. For the $2C_{H_{0/1}}$ SVM, we set $y_i = 1$ for the images from the manipulated class and $y_i = 0$ for the pristine ones. For the 1C SVMs ($1C_{H_0}$, $1C_{H_1}$,and $1C_{H_0}^{cmb}$) instead, the training is unlabeled, that is, the SVMs are trained with (${x_i}$) only, and the decision function is given by \eqref{dec_function} with  $y_i = 1$, $\forall i$ \cite{Scholkopf2000oneClassSVM2}.
In this work, we adopt an RBF (Radial Basis Function) kernel for the SVMs, that is, $K_j({x_i},x) = \exp ( - \gamma_j {\left\| {{x_i} - x} \right\|^2})$, where  $\gamma_j$ determines the width of the Gaussian kernel.

In a similar way, the decision function of the final combination classifier ($1C_{H_0}^{cmb}$) is expressed by,
\begin{equation}
\label{dec_function2}
f(x) = \sum\limits_{i = 1}^n {{\alpha_{f,i}}K_f({{\bf d}(x_i)},{\bf d}(x)) + b_f }
\end{equation}
where ${\bf d}(x) = (d_1(x), d_2(x), d_3(x))$ is the vector of the soft outputs of the intermediate classifiers (${\bf d}(x) \in \mathds{R}^3$) when the input feature vector is $x$, $0 \le \alpha_{f,i} \le C_f$ and $C_f$ is the cost term. The kernel $K_f$ is again an RBF with parameter $\gamma_f$. The best parameters $\gamma_j^*$ and $C_j^*$ (and $\gamma_f^*$ and $C_f^*$) of the classifiers, often referred to as hyper-parameters or internal parameters, are determined during the validation phase.

Some observations are in order.
In a 2C SVM, the parameter $C$ rules the tradeoff between the margin of the separating hyperplane in the higher dimensional space (the transformation of the input $x$ to the higher-dimensional space defines the kernel  \cite{boser1992training}) and the misclassification of the training points. A large $C$ means that getting all the training points classified correctly, and then a smaller margin, is preferable, even if this goes with the risk of data overfitting. For 1C SVMs, according to the formulation by Sch\"olkopf et al. \cite{Scholkopf2000oneClassSVM1},  the selection of the hyper-parameters is conventionally carried out by considering $\gamma$ and $\nu = 1/C$ (rather than  $\gamma$ and $C$), where $\nu$ determines the margin of the decision region in the higher-dimensional space. More specifically, the parameter $\nu$ sets an upper bound on the fraction of errors, i.e., training samples being misclassified \cite{Scholkopf2000oneClassSVM2} (for instance, by setting $\nu = 0.05$, at most 5\% of the training samples are allowed to be wrongly classified).

The most important parameter for both 2C and 1C SVMs in the case of  RBF kernel is $\gamma$, which determines the width of the kernel and then determines how far the influence of a training sample reaches. Specifically, $\gamma$ defines the inverse of the radius of influence: the smaller is $\gamma$, the fewer support vectors are selected and the decision region becomes more spherical. Essentially, $\gamma$ regulates the tradeoff between capturing the complex shape of the data (large $\gamma$) and avoiding overfitting (small $\gamma$).
If $\gamma$ is too large, the radius of the area of influence of the support vectors includes {\em almost only} the support vector itself and no choice of the regularization terms $C$ and $\nu$ is able to prevent overfitting.

\subsubsection{Choice of the hyper-parameters of the intermediate classifiers} 
We optimized the hyper-parameters of  $2C_{H_{0/1}}$  on the validation set by means of an exhaustive search. To do so, we first split the validation set into a training and test set. Then we trained the system for every choice of the parameters $(C, \gamma)$, then we chose the pair with the best test accuracy, that is the pair that minimizes $P_e = 0.5 P_{fa} + 0.5 P_{md}$.
For better performance in terms of generalization capability, standard $v$-fold cross validation was also performed for every $C$ and $\gamma$ \cite{montgomery2017design}, that is, we repeated the process $v$ times each time by splitting the set in a different way. The pair $(C, \gamma)$ with the best average cross-validation accuracy was then selected and used for training the system on the training set.

\begin{figure}
	\centering
	\subfigure[$2C_{H_{0/1}}$: $\alpha = \beta$]{\label{fig:a}\includegraphics[width=45mm]{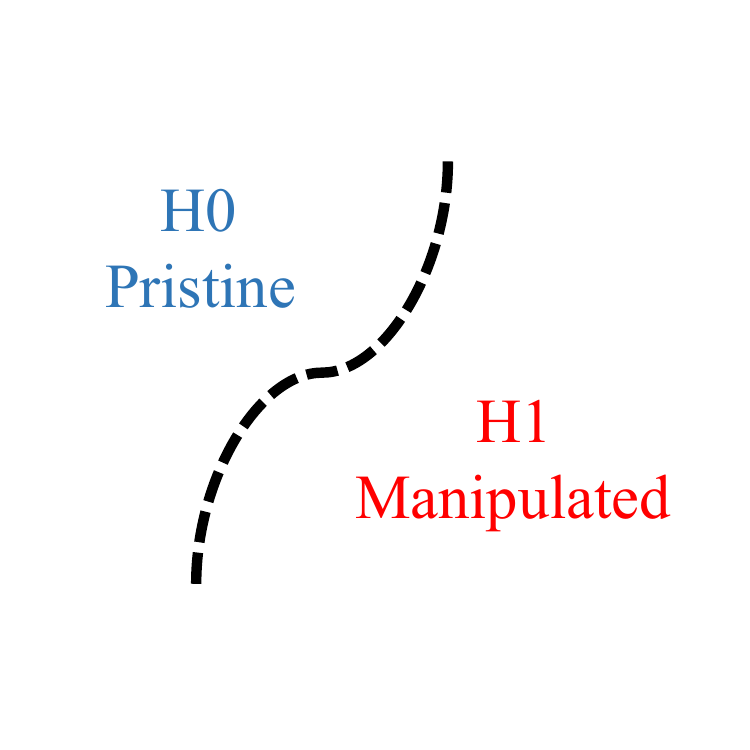}}
	\subfigure[$1C_{H_0}$: $\alpha < \beta$]{\label{fig:b}\includegraphics[width=45mm]{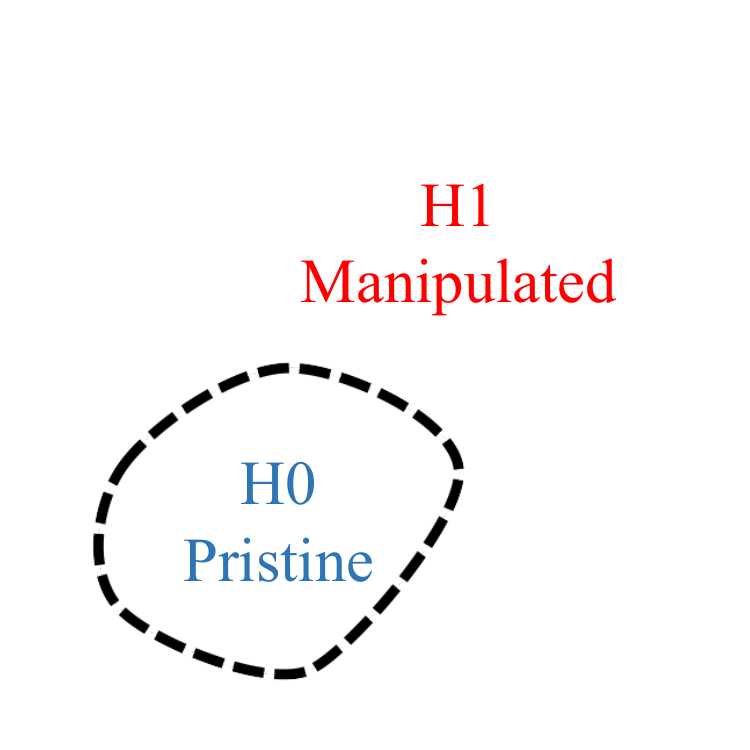}}
	\subfigure[$1C_{H_1}$: $\alpha < \beta$]{\label{fig:c}\includegraphics[width=45mm]{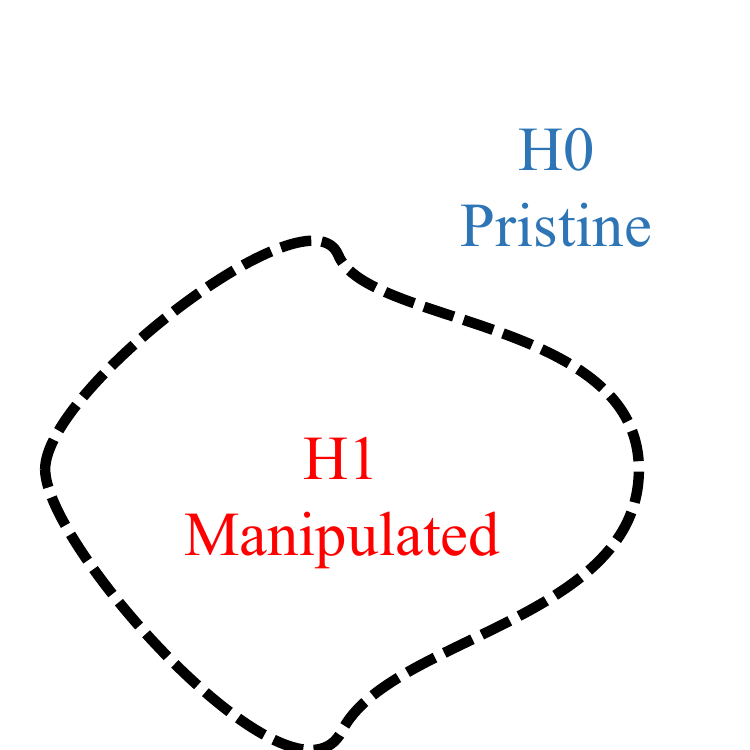}}
	\subfigure[$1C_{H_0}^{cmb}$: $\alpha < \beta$]{\label{fig:d}\includegraphics[width=45mm]{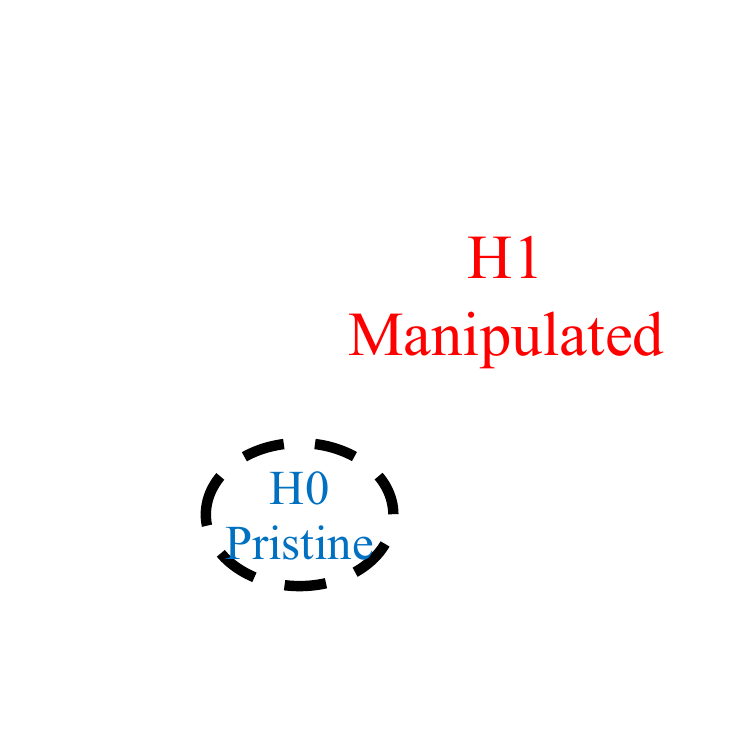}}
	\caption{Pictorial representation of the decision regions of the classifiers composing the overall 1.5C detector.
		To get an advantage in terms of security, the 1C classifiers are designed by letting $\alpha < \beta$. As a consequence, the 1C classifiers trained on pristine images - (b) and (d)- will have smaller acceptance regions, whereas the  1C  trained on manipulated images - (c) - will have quite a large acceptance region.}
	\label{fig:setup}
\end{figure}

Similarly, for $1C_{H_0}$ and $1C_{H_1}$, we carried out an exhaustive search over both $\gamma$ and $\nu$ to find the pair leading to the best accuracy. However, the 1C classifiers tend by construction to have poor performance with respect to the alternative class, i.e., the class of samples not used for training. Since  we wish to avoid missed detection events  ($H_1$ detected as $H_0$),  in order to increase the security against integrity violation attacks, we validated the 1C SVMs by weighting differently the two kinds of error probabilities.
Let $\alpha$ and $\beta$ be the weights assigned to the probability of a false alarm and a missed detection, respectively. While for $2C_{H_{0/1}}$ we let $\alpha = \beta$,  for the 1C classifiers we set $\alpha < \beta$ so that the classifiers are trained in such a way to minimize the error probability term $\alpha P_{fa} + \beta P_{md}$.  This corresponds to consider a {\em relatively small} closed acceptance region for  $1C_{H_0}$ and $1C_{H_0}^{cmb}$, and a {\em relatively large} closed region for $1C_{H_1}$. The situation is illustrated in Fig. \ref{fig:setup}. According to the adversarial setup considered in our work (see scheme of Fig. \ref{Fig1.setup}(b)), in fact, the attacker aims at entering the pristine region, or equivalently, exiting the manipulated region; then, choosing $\alpha$ lower than $\beta$ should improve the performance of the 1C classifiers in the presence of attacks. Obviously, a considerable unbalance between the two weights may imply worse performance in the absence of attacks. However, we verified that, thanks to the presence of $2C_{H_{0/1}}$, the overall robustness of the 1.5C classifier remains good even when $\alpha$ is much lower than $\beta$ and then the 1C SVMs are designed by focusing more on the security performance, at the possible cost of a lower robustness.

\section{Evaluation Methodology}
\label{sec.methodology}

The goal of the 1.5C detector is to combine the good performance of a 2C classifier in the absence of intentional attacks (robustness) and the improved security achieved by 1C classification (security). To prove that this is indeed the case, we will show that the 1.5C detector is more robust than the intermediate 1C classifiers and that attacking the 1.5C detector requires a larger distortion with respect to attacking the intermediate 2C classifier. In the next subsections, we describe the exact methodology we have followed for our tests, while the results we have got are presented and discussed in Section \ref{sec.results}.

\subsection{Goal of the Detectors}
\label{sec.processing}

\begin{figure}
	\centering
	\subfigure[Pristine image]{\label{fig1:a}\includegraphics[width=55mm]{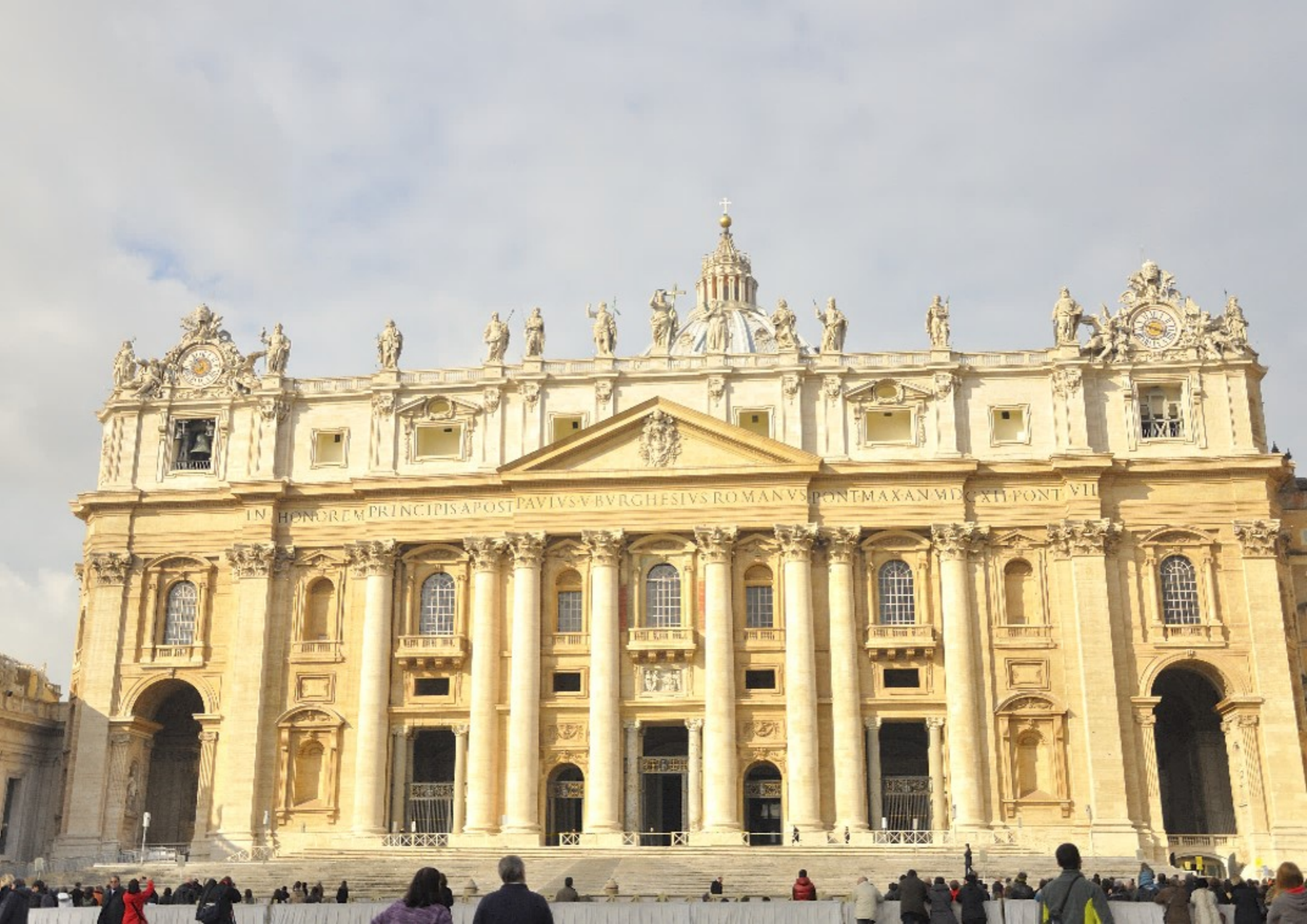}}
	\subfigure[Histogram equalized image (CL-AHE)]{\label{fig1:b}\includegraphics[width=55mm]{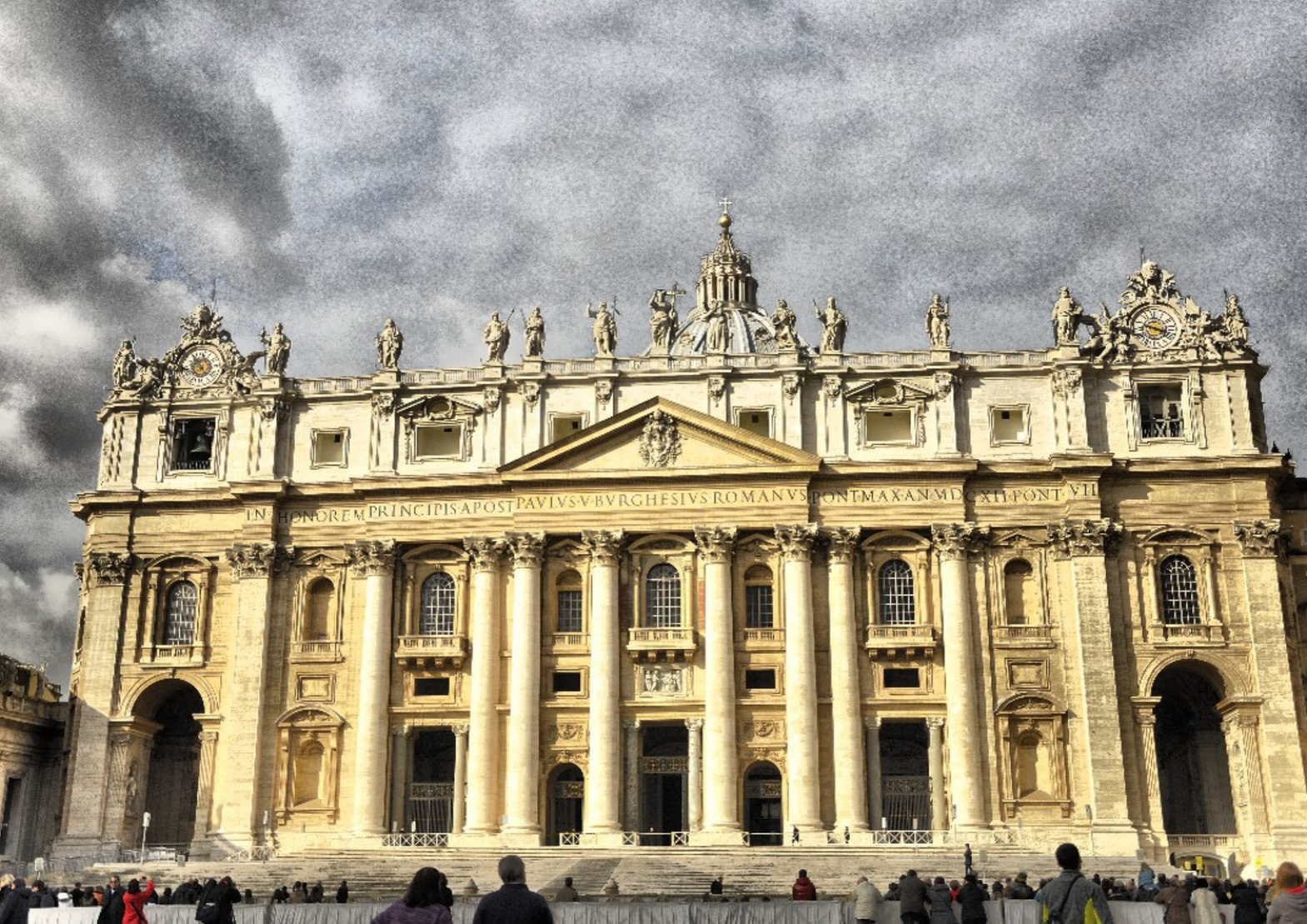}}
	\subfigure[Median filtered image]{\label{fig1:c}\includegraphics[width=55mm]{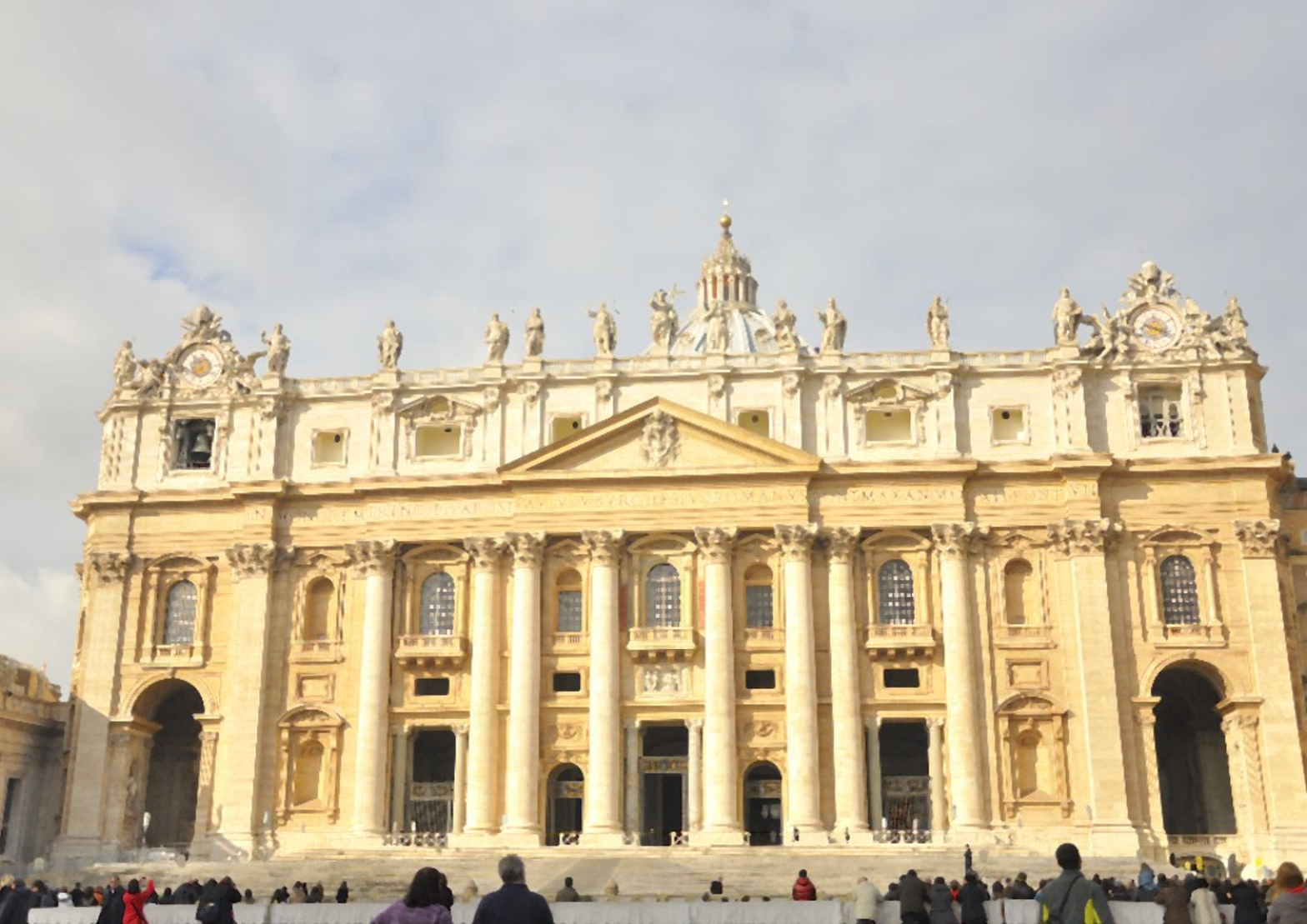}}
	\subfigure[Resized image]{\label{fig1:d}\includegraphics[width=75mm]{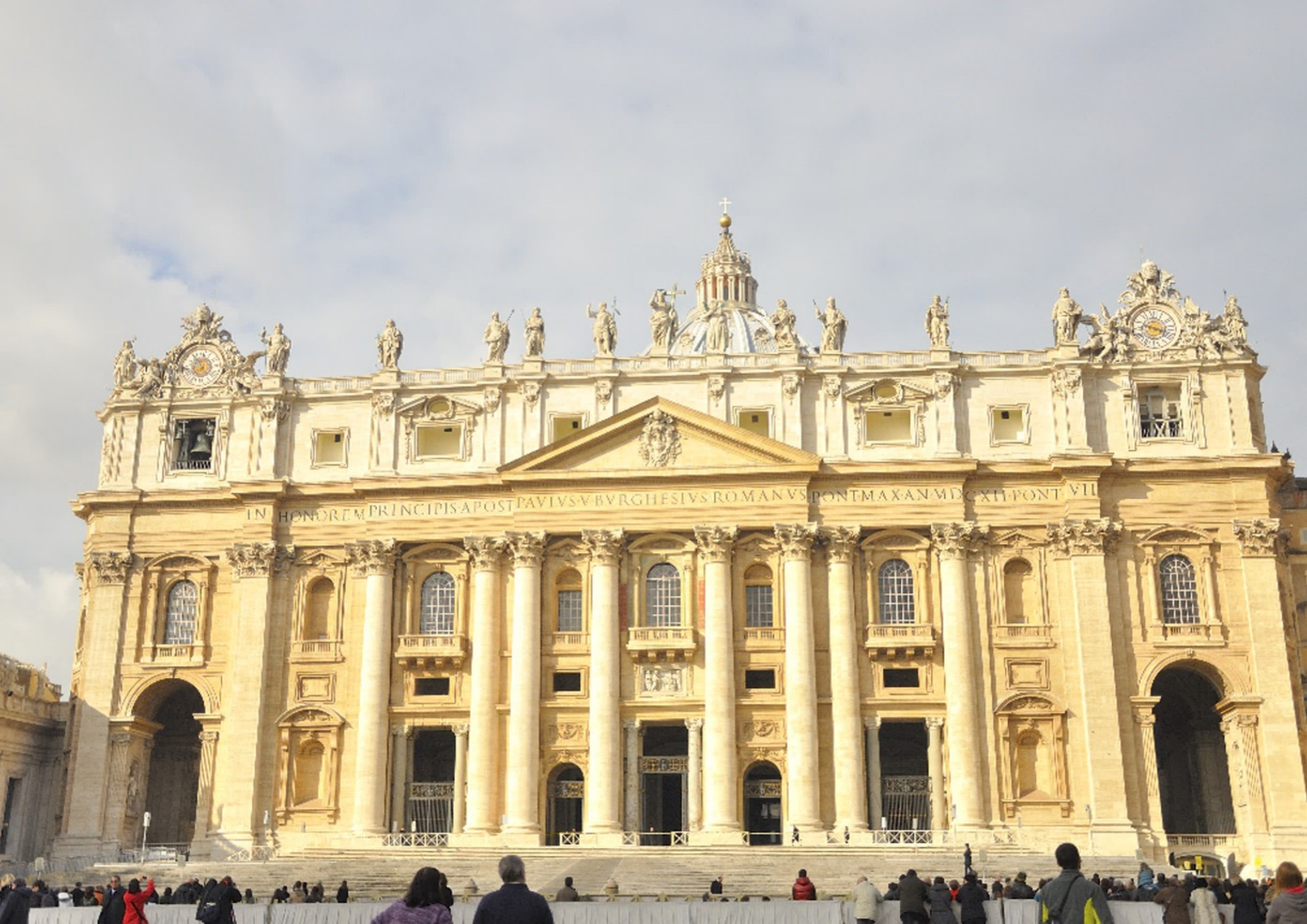}}
	\caption{A visual comparison between a pristine image ($H_0$) and its manipulated versions ($H_1$), for the processing operators considered in this paper. The clip-limit parameter for the CL-AHE is 0.05; the windows size for the median filter is $3\times 3$; resizing is applied with a zooming factor equal to 1.3.
	}
	\label{fig1:setup}
\end{figure}

To asses the effectiveness of the 1.5C architecture, we focused on the detection of three different kinds of image manipulations, namely geometric
transformation, filtering and contrast enhancement. Specifically, we considered the following three processing operations: resizing, median filtering and histogram equalization. With regard to resizing, we considered a bicubic interpolation and a resizing scaling factor (zooming factor) equal to 1.3.
For median filtering,  we set the window size to $3 \times 3$, so to keep the visual degradation of the filtered image limited.  Finally, for histogram equalization, we considered the Clip-Limited Adaptive Histogram Equalization (CL-AHE) algorithm \cite{Zuiderveld1994}.
With respect to standard Adaptive Histogram Equalization (AHE),  CL-AHE does not over-amplify noise  in relatively homogeneous regions as done by AHE, by clipping the  histogram before computing the enhancement transformation. In our experiments, the clip-limit parameter was set to 0.05.
When working with color images, the CL-AHE operator is applied  to the luminance channel, precisely to the V channel (then, the image is  converted from the RGB to the HSV color space). This is a commonly adopted strategy, since the straightforward application of CL-AHE to each color channel separately would unnaturally change the color balance and produce a visually unpleasant image.
The same strategy is followed for the case of median filtering, which, once again, is applied to the luminance channel only (the V channel in our case). An example of a pristine image and the corresponding processed images is given in Fig. \ref{fig1:setup}.
\subsection{Datasets Creation}
\begin{figure}[t!]
	\centering
	\includegraphics[width=0.8\textwidth , height = 3cm]{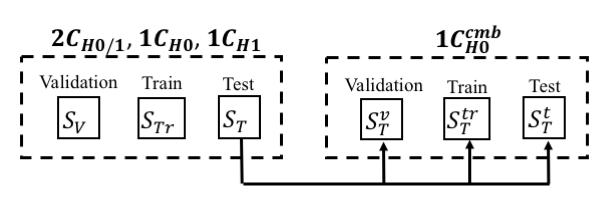}
	\caption{Datasets used for training and testing the classifiers of the 1.5C system. $\{S_T^v \cup S_T^{tr} \cup S_{T}^{t}\} \equiv S_T$.}
	\label{Fig5.setup}
\end{figure}

To produce the datasets for our experiments, we considered camera-native (uncompressed) images. The images of the dataset were used to training, validate and test the three intermediate classifiers ($2C_{H_{0/1}}$, $1C_{H_0}$,and $1C_{H_1}$). The set of images used for testing is further split to build the training and test set for the final combination classifier $1C_{H_0}^{cmb}$.
More specifically,  let us denote with $S_V$ the set of images used for the internal validation of the hyper-parameters of the intermediate classifiers, with $S_{Tr}$ the set used for training, and with $S_T$ the  set used for testing, see Fig. \ref{Fig5.setup}. The test set $S_T$ is further split into three sets, namely $S_T^{v}$, $S_T^{tr}$ and $S_T^{t}$,  used, respectively, for internal validation, training  and testing of $1C_{H_0}^{cmb}$. Since the dimensionality of the input feature vector of the downstream 1C is very low, corresponding to the three soft outputs of the intermediate classifiers (${\bf d} \in \mathds{R}^3$), the number of images in $S_T^{v}$, $S_T^{tr}$ and $S_T^{t}$ (and hence in $S_T$) does not need to be very large.

Starting from the above sets of images, no processing is applied to build the samples of the first class ($H_0$ - pristine images), see Fig. \ref{Fig1.setup}(a). For the second class ($H_1$), the samples are built by applying different processing operators, as detailed in Section \ref{sec.processing}.
\subsection{Robustness Analysis}

The robustness of the 1.5C detector is assessed by adopting the setup illustrated in Fig. \ref{Fig1.setup}(a). Therefore, the system is trained and tested under the same conditions, i.e., by assuming that there is no attack at test time.

The performance of the 1.5C system are measured over the test set $S_T^t$, i.e., the set used for testing the final 1C classifiers. The metric used is the Area Under Curve (AUC) of the ROC curve of the classifier. By comparing these results with those achieved by $2C_{H_{0/1}}$ (which is tested on the entire $S_T$ set - see Fig. \ref{Fig5.setup}), we can compare the performance of the 1.5C architecture with respect to a conventional 2C classifier, and assess the - possible -  drop of performance experienced in the former case.\footnote{We  verified experimentally that the fact that the performance of 2C and 1.5C are not tested exactly on the same set does not affect the results.}
In general, 1C classifiers are known to have poor robustness in presence of post-processing, limiting their applicability in practice.
To show that this is not the case with the overall 1.5C architecture,
we also run some experiments to assess the robustness performance of the 1.5C classifier in the presence of noise addition and JPEG post-processing.
Results confirm that, thanks to the presence of the intermediate 2C classifier, the 1.5C system performs much better than the intermediate 1C detectors in terms of robustness. For sake of simplicity, for these tests, we only considered the case of resizing detection. The exact results of our experiments will be detailed in Section \ref{sec.results}.

\subsection{Security Assessment}
The security of the 1.5C classifier is assessed by evaluating the performance of the system under attacks (see Fig. \ref{Fig1.setup}(b)). These performance are compared against those achieved by $2C_{H_{0/1}}$, when tested under the same attack.
The goal of this analysis is to validate the expectation that the 1.5C architecture offers a security advantage over 2C classification, in that attacking the system introduces a larger distortion into the attacked images.

To be more specific, we considered the following attack model:\\

\textbf{Attacker's Goal}: the attacker wants to modify a manipulated  ($H_1$) image in order to induce a missed detection event, i.e., in such a way to move the feature vector of the image into the region of pristine images ($H_0$ region).\\

\textbf{Attacker's knowledge}:
by relying on the terminology in \cite{Biggio2013evasionAttack}, we focus on a Perfect Knowledge (PK) scenario, in which the attacker has a complete knowledge about the classifier. This corresponds to consider the most favourable case for the attacker.\\

\textbf{Attacker's capability}:
We focus on exploratory attacks \cite{huang2011adversarial}, that is, attacks carried out at test time. The large majority of the counter-forensic methods proposed in the literature belong to this category.\\

In the above scenario, the attacker aims at inducing a missed detection error, by minimizing the distortion introduced into the image. It is worth stressing that, a targeted PK attack is always successful in causing a misclassification, i.e., it always enter the $H_0$ region. However, it is expected that attacking a more secure classifier will require a larger distortion.

Regarding the attack algorithm, we considered the gradient-based attack against SVM detectors described in \cite{Chen2017GraAttack}.
The attack works by computing an approximation of the gradient of the SVM output with respect to the image pixels. The approximated gradient provides an approximation of the steepest descent direction of the decision function. Then, the step size (strength of the attack) is adjusted by controlling the percentage of modified pixels. If the image cannot be successfully attacked by modifying a maximum prescribed fraction of pixels, the modification is applied and the process is iterated.
As opposed to other approaches, the attack proposed in \cite{Chen2017GraAttack} is directly carried out in the pixel domain and then it can be applied even when the relationship between the feature and the pixel domain is not invertible, as it is the case with the SPAM features considered in this paper. The implementation of the attack passes through the definition of a safety margin $\rho$ \cite{Chen2017GraAttack}, which determines how much the attack goes inside the acceptance region. By choosing a larger $\rho$ (so to move the attacked image more deeply inside the $H_0$ region), the attack is more robust to perturbations of the decision boundary. This advantage goes at the price of a larger distortion introduced in the image

In order to compare the security of the 1.5C classifier with respect to the 2C one, we run the targeted attack in \cite{Chen2017GraAttack} against $2C_{H_{0/1}}$ and $1C_{H_{0}}^{cmb}$.
The performance of the classifiers under attacks are assessed on a subset of images in $S_{T}^t$, processed with median filtering, resizing and CL-AHE and then attacked by means of the attack described above.

\section{Experimental Results}
\label{sec.results}

The camera-native (uncompressed) images used for our experiments were taken from the RAISE-8K dataset \cite{RAISE8K}. Specifically, a total amount of 7997 images were used, split as follows (see Fig. \ref{Fig5.setup}): 1000 images were selected to build the validation set $\mathcal{S}_V$, 5000 for the training set $\mathcal{S}_{Tr}$, and the remaining 1997 for $\mathcal{S}_T$. Then, the images in $\mathcal{S}_T$ were further split to build the validation, training and test sets used for the combination classifier as follows: 300 images were used to build $\mathcal{S}_T^v$, 700 for $\mathcal{S}_T^{tr}$ and the remaining 997 images to build $\mathcal{S}_T^t$. Note that a much lower number of images would be sufficient for testing the final SVM which is trained on just 3-dimensional input feature vectors.
The images from every set were then processed to build the class of  $H_1$ samples, whereas the unprocessed images were used to build the class of pristine images ($H_0$).
For security assessment, the attack in  \cite{Chen2017GraAttack} was applied to 100 images in the set $\mathcal{S}_T^t$ belonging to the $H_1$ class. To speed up the feature extraction step and the attacks, we sub-sampled the images from the RAISE8K dataset down to a size of $1072 \times 770$ without interpolation.

The Matlab environment was used to process the images and to design the classifiers of the 1.5C system. All SVMs were trained and tested by using the LibSVM library package \cite{LIBSVM}. We run our experiments on a system hardware Intel(R) Core i7-6700 CPU @ 3.40 GHz with four cores, and 32 GB of RAM.

\subsection{Hyper-parameters Setting}

As anticipated in Section \ref{sec.featureSelection}, the SPAM features are extracted from the V channel, obtained by converting the image from the RGB to the HSV color space.

Regarding the weights assigned to the two types of error probabilities during the validations phase, we set $\alpha=0.2$ ($\beta = 0.8$) for $1C_{H_0}$ and  $1C_{H_1}$ and $\alpha=0.1$ ($\beta = 0.9$) for $1C_{H_0}^{cmb}$. In making this choice of $\alpha$ and $\beta$, we verified that the system robustness is not affected too much by the use of heavily asymmetric weights.
For validating the internal parameters of $2C_{H_{0/1}}$, we followed the standard exhaustive search method (known as grid-search) in the LibSVM  library \cite{LIBSVM}, which considers exponentially growing values for $C$ and $\gamma$ to identify the best parameters \cite{hsu2003practical}. Specifically, we considered the following grid-search area: $C \in \{ {2^{ -5}},{2^{ - 3}},...,{2^{15}}\}$ and $\gamma \in \{ {2^{ - 15}},{2^{ - 13}},...,{2^{3}}\}$ and performed a $5$-fold cross validation (i.e. $\nu = 5$).

To set the hyper-parameters of $1C_{H_0}$, $1C_{H_1}$, and $1C_{H_0}^{cmb}$, we followed a similar strategy by taking $\nu $, $\gamma  \in \{ {2^{ - 10}},{2^{ - 9}},...,{2^9},{2^{10}}\}$.
In addition, since the distribution of the samples used for internal parameter validation is very important to learn correctly the hyper-parameters of the 1-C SVMs, we used the entire training set to train the SVMs during the exhaustive search; then, the validation set was used only to perform testing in this phase and choose the pair ($\gamma,\nu$) providing the best accuracy. To limit the computational burden, $v$-fold cross validation was not performed in this case.

Table \ref{tab:HyperSVMs} shows the best hyper-parameters ($C^*$, $\gamma^*$) for $2C_{H_{0/1}}$  and  ($\nu^*$,$\gamma^*$) for the three 1C SVMs.
From the tables, we see that, for the $1C_{H_0}^{cmb}$ SVM, the minimum values for the $\nu$ and $\gamma$ parameters are selected,
meaning that the SVM is able to get a low probability of erroneous classification of the training samples by relying on very few support vectors, hence minimizing the risk of overfitting.

\begin{table}
	\caption{Hyper-parameters of the SVMs classifiers.}
	\label{tab:HyperSVMs}      
	\begin{tabular}{lllll}
		\hline\noalign{\smallskip}
		&   {\bf $2C_{H_{0/1}}$} & {\bf $1C_{H_0}$} & {\bf $1C_{H_1}$} & {\bf $1C_{H_0}^{cmb}$} \\
		\noalign{\smallskip}\hline\noalign{\smallskip}
		\multirow{2}{*}{Resizing} & $C^* = 2^{11}$ & $\nu^* = {2}^{-3}$ & $\nu^* = {2}^{-9}$ & $\nu^* = {2}^{-10}$   \\
		& $\gamma^* = 2^{-1}$ & $\gamma^* = {2}^{9}$ & $\gamma^* = {2}^{7}$ & $\gamma^* = {2}^{-10}$  \\
		\noalign{\smallskip}\hline\noalign{\smallskip}
		\multirow{2}{*}{Median filter} & $C^* = 2^{11}$ & $\nu^* = {2}^{-5}$ & $\nu^* = {2}^{-9}$ & $\nu^* = {2}^{-10}$   \\
		& $\gamma^* = 2^{-1}$ & $\gamma^* = {2}^{7}$ & $\gamma^* = {2}^{6}$ & $\gamma^* = {2}^{-10}$  \\
		\noalign{\smallskip}\hline\noalign{\smallskip}
		\multirow{2}{*}{CL-AHE} & $C^* = 2^{11}$ & $\nu^* = {2}^{-3}$ & $\nu^* = {2}^{-10}$ & $\nu^* = {2}^{-10}$   \\
		& $\gamma^* = 2^{-1}$ & $\gamma^* = {2}^{9}$ & $\gamma^* = {2}^{7}$ & $\gamma^* = {2}^{-10}$  \\
		\noalign{\smallskip}\hline
	\end{tabular}
\end{table}

\subsection{Performance in the Absence of Attacks}
\begin{figure}
	\centering     
	\subfigure[$2C_{H_{0/1}}$]{\label{fig20:a}\includegraphics[width=0.70\columnwidth]{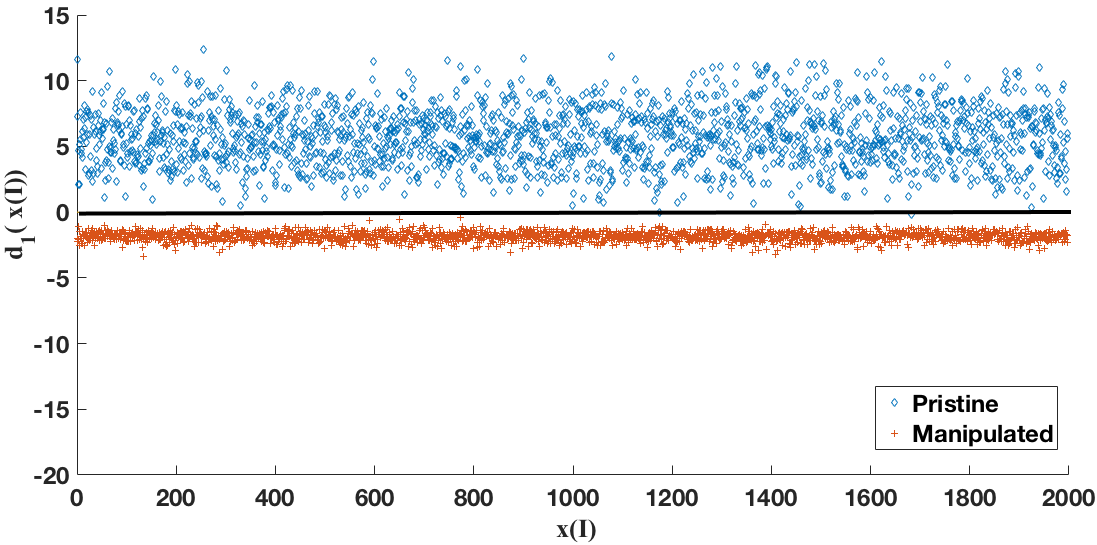}}
	\subfigure[$1C_{H_0}$]{\label{fig20:b}\includegraphics[width=0.70\columnwidth]{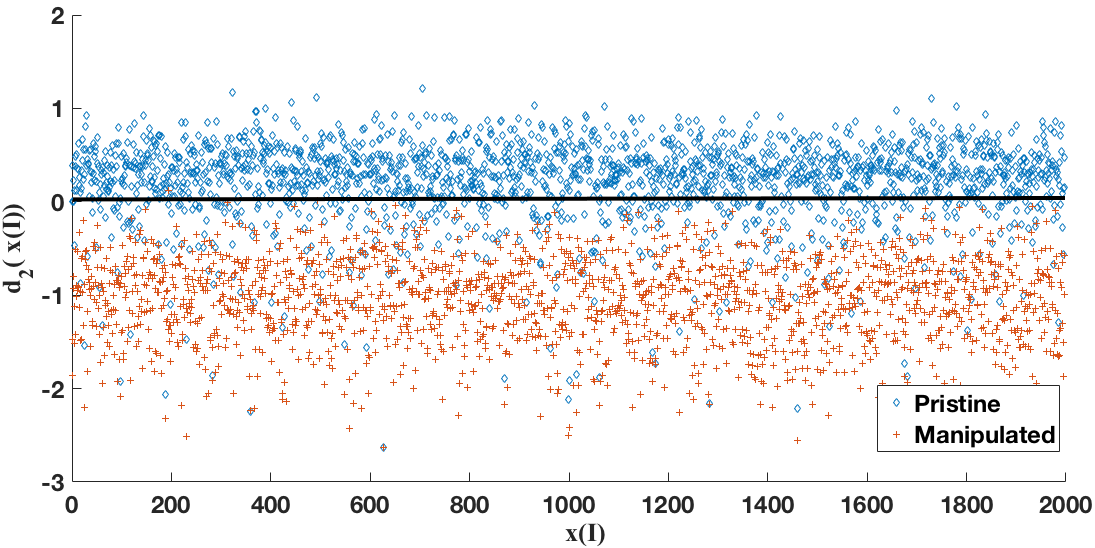}}
	\subfigure[$1C_{H_1}$]{\label{fig20:c}\includegraphics[width=0.70\columnwidth]{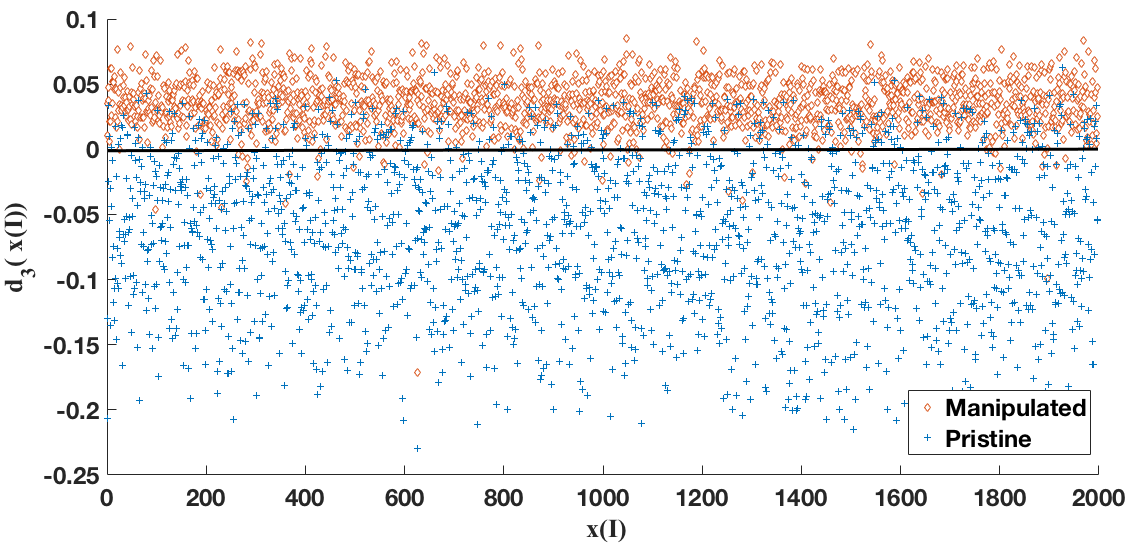}}
	\subfigure[$1C_{H_0}^{cmb}$]{\label{fig20:d}\includegraphics[width=0.70\columnwidth]{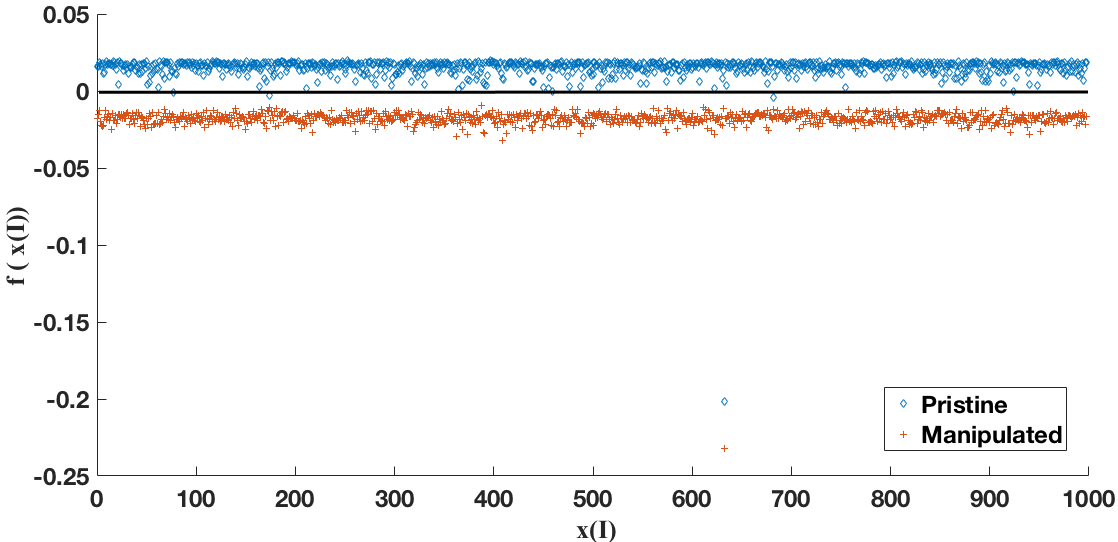}}
	\caption{Decision values of the four SVMs trained for resizing detection on the test set, for both $H_0$ (pristine samples) and $H_1$ (manipulated samples). Decision values of $2C_{H_{0/1}}$ for the images in $S_T$ (a); decision values of the intermediate 1C classifiers ($1C_{H_0}$,and $1C_{H_1}$)  for the images in $S_T$ (b)-(c); decision values of $1C_{H_0}^{cmb}$ for the images in $S_T^t$ (d).}
	\label{fig20:setup}
\end{figure}

The values of the decision functions of the four SVM classifiers over the test set are reported in Fig. \ref{fig20:setup} for the case of resizing detection.
We see that $2C_{H_{0/1}}$ is able to tell apart pristine and manipulated images, obtaining perfect classification in the absence of attacks; moreover, the scatter plot shows that the clouds of points are very well separated. The two intermediate 1C SVMs also achieve high-accuracy, but the classification is not perfect. Finally, $1C_{H_{0}}^{cmb}$  achieves almost perfect classification, similarly to $2C_{H_{0/1}}$. For both $2C_{H_{0/1}}$ and $1C_{H_{0}}^{cmb}$, the decision threshold is set to 0. Note that while for  $2C_{H_{0/1}}$,  $1C_{H_{0}}$ and $1C_{H_{0}}^{cmb}$,  the label $y=1$ is assigned to the images of the pristine class ($H_0$), for  $1C_{H_{1}}$, $y=1$ is assigned to the manipulated class ($H_1$), and that is why the scatter plots in Fig. \ref{fig20:c} are reverted.

Very similar results were obtained for median filtering and  CL-AHE. Table \ref{tab:AUCnoAttack} shows the AUC values of the ROC curve for $2C_{H_{0/1}}$  and the 1.5C classifiers, as well as those of the intermediate 1C SVMs, for each detection task. We observe that by using $1C_{H_{0}}^{cmb}$, instead of $2C_{H_{0/1}}$, the performance drops very slightly in all the cases.
\begin{table}
	\caption{AUC values of all the classifiers for the three manipulation detection tasks. The performance of the 1.5C system are those reported for $1C_{H_0}^{cmb}$.}
	\label{tab:AUCnoAttack}     
	\begin{tabular}{lllll}
		\hline\noalign{\smallskip}
		& {\bf $2C_{H_{0/1}}$} & $1C_{H_0}$ & $1C_{H_1}$ & {\bf $1C_{H_0}^{cmb}$}  \\
		\noalign{\smallskip}\hline\noalign{\smallskip}
		Resizing & {\bf 1} & 0.96 & 0.96 & {\bf 0.99} \\
		Median Filter & {\bf 1} & 0.99 & 0.99 & {\bf 0.99}  \\
		CL-AHE & {\bf 1} & 0.95 & 0.97 & {\bf 0.99}\\
		\noalign{\smallskip}\hline
	\end{tabular}
\end{table}

\subsubsection{Robustness of the Classifiers}

To assess the robustness of the 1.5C architecture compared to the 2C and the 1C detectors, we evaluated the performance of $2C_{H_{0/1}}$, $1C_{H_0}$, $1C_{H_1}$ and  $1C_{H_0}^{cmb}$ in the presence of Gaussian noise with zero mean and variance $\sigma^2 = 5\cdot 10^{-6}$, $10^{-5}$, $1.5 \cdot 10^{-5}$ and $2\cdot 10^{-5}$ (standard deviation ranging from $\sigma = 0.0022$ to $\sigma = 0.0045$),
and in the presence of JPEG compression with Quality Factors (QF) 85, 90, 95, and 98. For the case of noise addition, the average Mean Square Error (MSE) introduced by the noise
ranges from  0.3 to 1.

Tables \ref{tab:ResizeHS_JPEG} and \ref{tab:ResizeHS_Noise} show the average accuracy of the tests on noisy images and JPEG compressed images respectively, for the resizing detection task.  We see that while the performance of the 1Cs classifiers are significantly impaired by the post-processing, the 1.5C classifier is more robust and its performance remain comparable to those of the 2C detector. 

Expectedly, if we consider a much larger noise (or a stronger compression), the performance of the classifiers drop. In order to design a classifier that works properly under these conditions, a possibility is to consider an aware classifier, which takes into account the possible presence of post-processing during the training phase, by including post-processed samples in the training set \cite{CFharder16,EUSIPCO17}. This analysis is outside the scope of this paper, since here we are interested in validating the 1.5C architecture, so we leave it for a future work.

\begin{table}
	\caption{Robustness of the classifiers in the presence of JPEG compression (accuracy).}
	\label{tab:ResizeHS_JPEG}     
	\begin{tabular}{lllll}
		\hline\noalign{\smallskip}
		QF & $2C_{H_{0/1}}$ & $1C_{H_0}$ & $1C_{H_1}$ & $1C_{H_0}^{cmb}$  \\
		\noalign{\smallskip}\hline\noalign{\smallskip}
		85  & 0.90 & 0.71 & 0.83 & 0.88 \\
		90  & 0.94 & 0.75 & 0.87 & 0.93 \\
		95  & 0.98 & 0.82 & 0.90 & 0.97 \\
		98  & 0.99 & 0.87 & 0.91 & 0.99  \\
		\noalign{\smallskip}\hline
	\end{tabular}
\end{table}
\begin{table}[h!]
	\caption{Robustness of the classifiers under noise addition (accuracy).}
	\label{tab:ResizeHS_Noise}     
	\begin{tabular}{lllll}
		\hline\noalign{\smallskip}
		Noise parameter & $2C_{H_{0/1}}$ & $1C_{H_0}$ & $1C_{H_1}$ & $1C_{H_0}^{cmb}$  \\
		\noalign{\smallskip}\hline\noalign{\smallskip}
		$5 \cdot 10^{-6}$  & 0.87 & 0.78 & 0.84 & 0.93 \\
		$10^{-5}$   & 0.84 & 0.79 & 0.85 & 0.92  \\
		$1.5 \cdot 10^{-5}$  & 0.79 & 0.80 & 0.87 & 0.89 \\
		$2 \cdot 10^{-5}$ & 0.74 & 0.82 & 0.89 & 0.83 \\
		\noalign{\smallskip}\hline
	\end{tabular}
\end{table}

\subsection{Performance under Attacks}

In this section, we assess the performance of $2C_{H_{0/1}}$ and the 1.5C classifiers in the presence of attacks \cite{Chen2017GraAttack}.
In all the experiments, the safety margin $\rho$ for the attack is set to 0.

\subsubsection{Attack against $2C_{H_{0/1}}$}

\begin{figure}
	\centering     
	\subfigure[$2C_{H_{0/1}}$]{\label{fig30:a}\includegraphics[width=0.70\columnwidth]{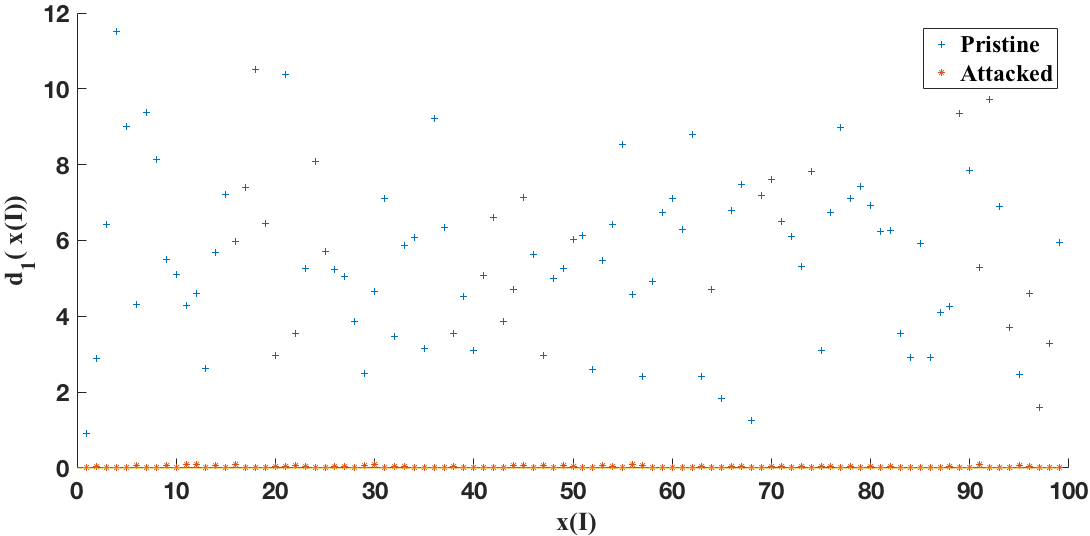}}
	\subfigure[$1C_{H_0}$]{\label{fig30:b}\includegraphics[width=0.70\columnwidth]{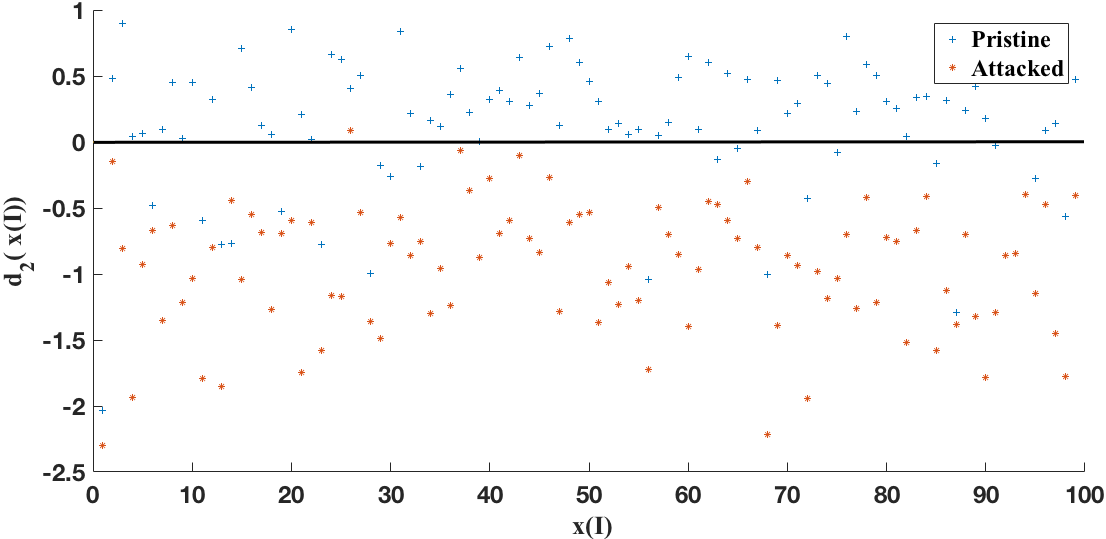}}\\
	\subfigure[$1C_{H_1}$]{\label{fig30:c}\includegraphics[width=0.70\columnwidth]{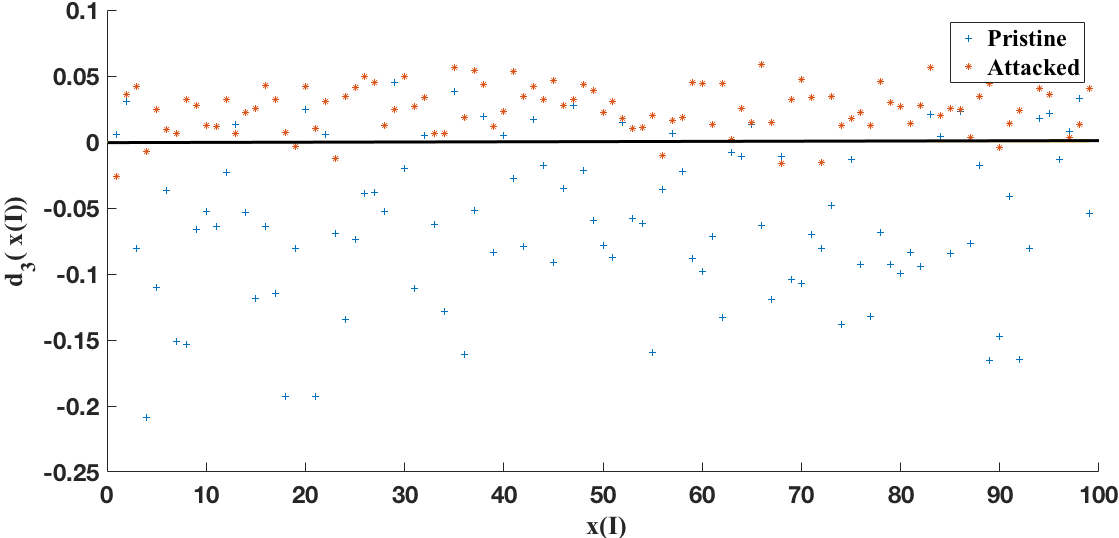}}
	\subfigure[$1C_{H_0}^{cmb}$]{\label{fig30:d}\includegraphics[width=0.70\columnwidth]{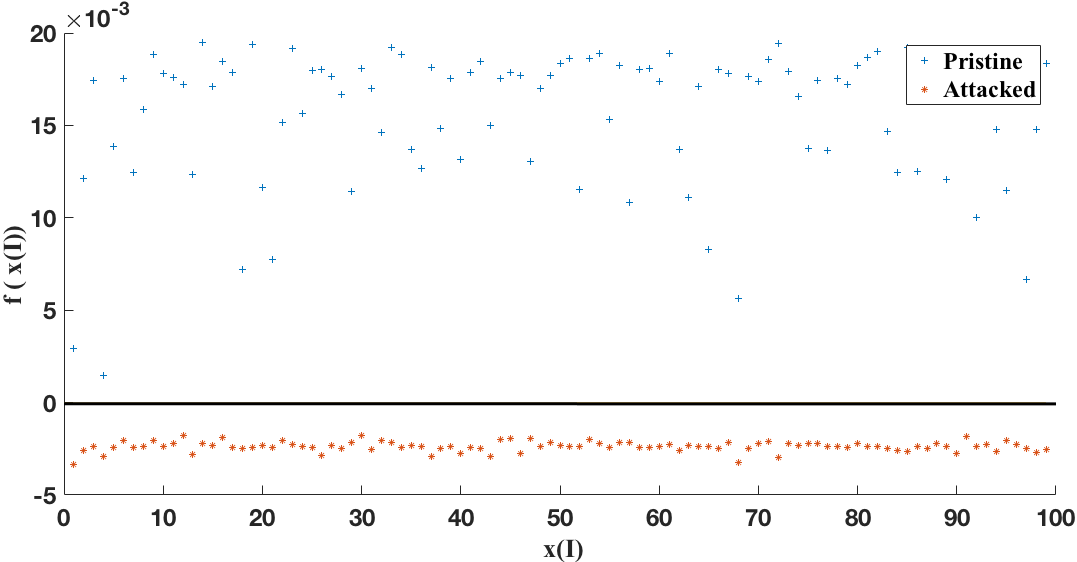}}
	\caption{Decision values of the four SVMs on the 100 images in $S_T^t$ in the presence of the attack in \cite{Chen2017GraAttack} under $H_1$, for the resizing detection task. The attack is carried out against the $2C_{H_{0/1}}$. Decision values of $2C_{H_{0/1}}$ (a); decision values of the intermediate 1C classifiers ($1C_{H_0}$, and $1C_{H_1}$) (b)-(c);  decision values of $1C_{H_0}^{cmb}$ (d).}
	\label{fig30:setup}
\end{figure}

For each detection task, we first run the attack against $2C_{H_{0/1}}$. As expected, the attack is always successful in inducing an incorrect classification and 100\% of the manipulated images are classified as pristine images after the attack. Moreover, all the images can be attacked in just one iteration of the algorithm in \cite{Chen2017GraAttack}. Fig. \ref{fig30:a} shows the results for the case of resizing detection.
Since we set $\rho = 0$, the attack stops as soon as the decision boundary is crossed.

The values of the decision function for the other SVMs of the 1.5C c classifier are shown in Fig. \ref{fig30:setup} (from \ref{fig30:b}  to \ref{fig30:d})\footnote{Note that, while for $2C_{H_{0/1}}$, $1C_{H_0}$ and $1C_{H_0}^{cmb}$ the attack is successful when it brings the pristine samples above the threshold, for $1C_{H_1}$, the goal of the attacker is to move the samples below the threshold.}.
From Fig. \ref{fig30:d}, we see that attacking $2C_{H_{0/1}}$  is not enough to fool the 1.5C classifier: the attacked samples in fact remain much below the decision threshold and the attack success rate is 0\%.
The success rate of all the attacks is reported in Table \ref{tab:2C_Attack}, where the percentage of misclassified attacked samples for the four SVMs is provided for the three detection tasks.
\begin{table}
	\caption{Percentage of misclassified attacked images. The attack is carried out against  $2C_{H_{0/1}}$.}
	\label{tab:2C_Attack}   
	\begin{tabular}{lllll}
		\hline\noalign{\smallskip}
		& {\bf $2C_{H_{0/1}}$} & $1C_{H_0}$ & $1C_{H_1}$ & {\bf $1C_{H_0}^{cmb}$}  \\
		\noalign{\smallskip}\hline\noalign{\smallskip}
		Resizing & {\bf 100\%} & 1\% & \%8 & {\bf 0\%} \\
		Median Filter & {\bf 100\%} & 4\% & 3\% & {\bf 0\%}  \\
		CL-AHE & {\bf 100\%} & 20\% & 12\% & {\bf 0\%}\\
		\noalign{\smallskip}\hline
	\end{tabular}
\end{table}
\subsubsection{Attack against the 1.5C classifier}

\begin{figure}[htbp]
	\centering     
	\subfigure[$2C_{H_{0/1}}$]{\label{fig40:a}\includegraphics[width=0.70\columnwidth]{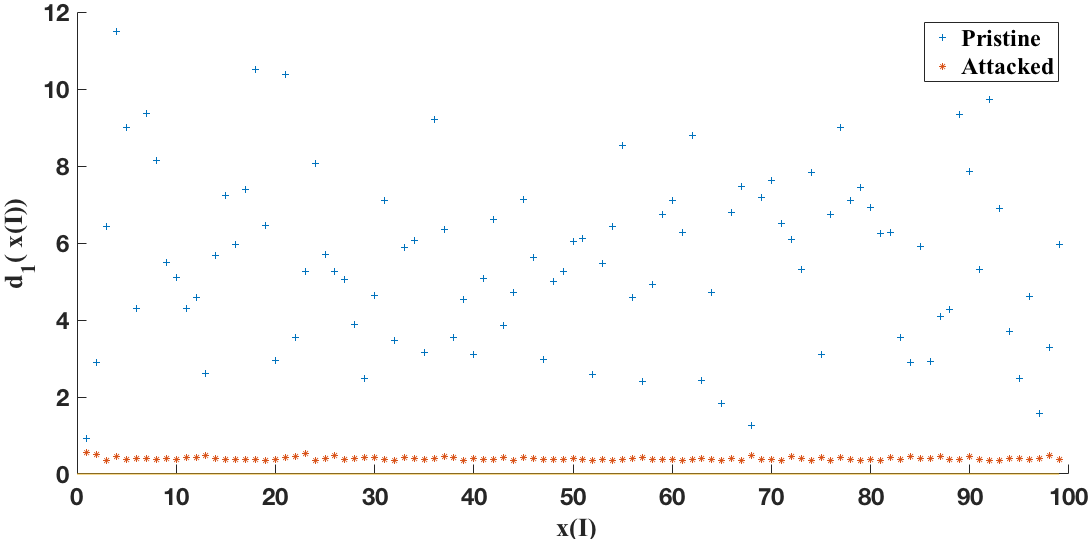}}
	\subfigure[$1C_{H_0}$]{\label{fig40:b}\includegraphics[width=0.70\columnwidth]{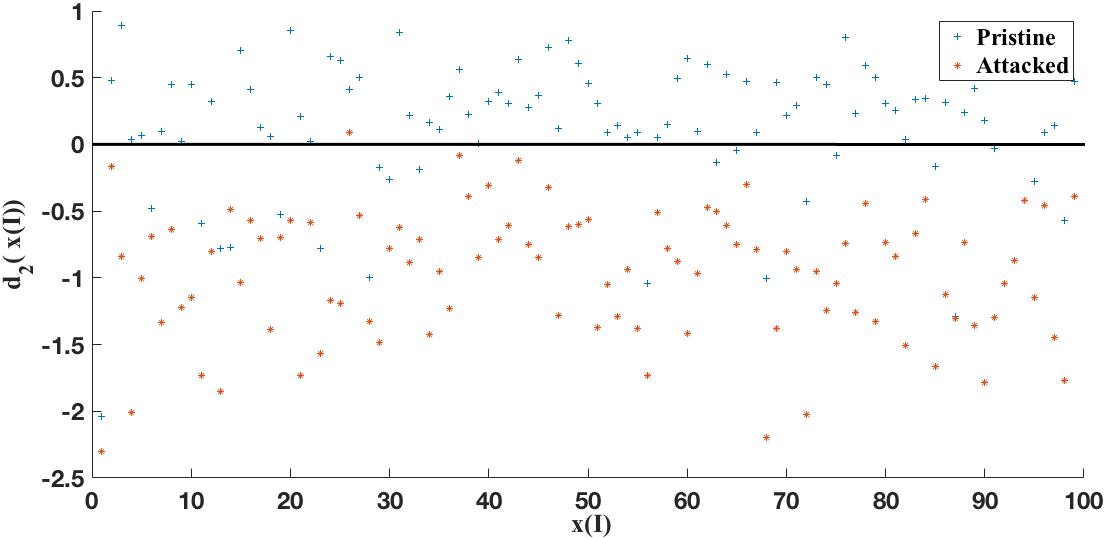}}
	\subfigure[$1C_{H_1}$]{\label{fig40:c}\includegraphics[width=0.70\columnwidth]{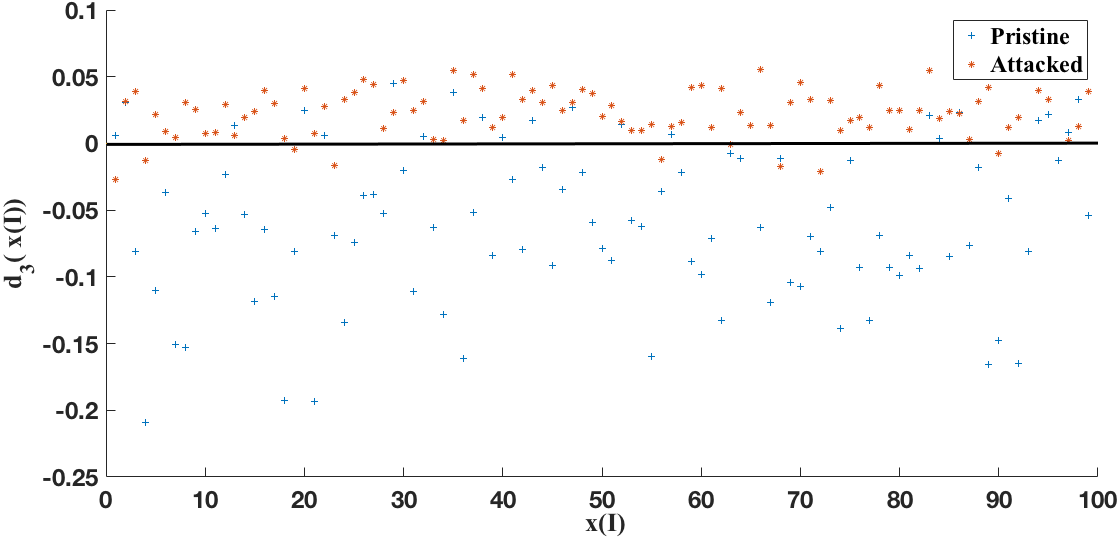}}
	\subfigure[$1C_{H_0}^{cmb}$]{\label{fig40:d}\includegraphics[width=0.70\columnwidth]{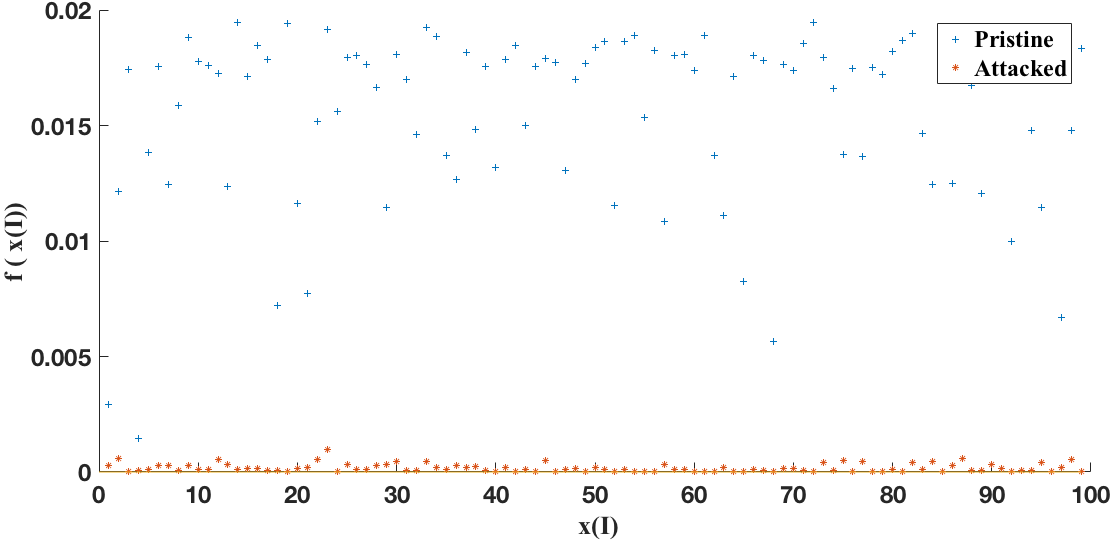}}
	\caption{Decision values of the four SVMs on the 100 images in $S_T^t$ in the presence of the attack in \cite{Chen2017GraAttack} under $H_1$, for the resizing detection task. The attack is carried out against the 1.5C classifier. Decision values of $2C_{H_{0/1}}$ (a); decision values of the intermediate 1C classifiers ($1C_{H_0}$,and $1C_{H_1}$) (b)-(c);  decision values of  $1C_{H_0}^{cmb}$ (d).}
	\label{fig40:setup}
\end{figure}

Figs. \ref{fig40:setup} shows what happens when the attack is carried out against the 1.5C classifier. In this case, most of the times, the attack requires more than one iteration to enter the $H_0$ region.
The Fig. refers to the case of resizing detection, however, similar results are obtained for the other manipulations. We observe that the values of the decision function for $2C_{H_{0/1}}$  on the attacked samples lie above the 0 threshold, and then the attack against the 1.5C is also effective against $2C_{H_{0/1}}$. Moreover, we see that the attack is not much effective against $1C_{H_0}$, and quite ineffective against $1C_{H_1}$, thus confirming that, thanks to the adoption of a closed acceptance region, the 1C classifiers are more difficult to attack.
Accordingly, the attack is successful in inducing a wrong classification for the 1.5C, mainly because $2C_{H_{0/1}}$ fails to a strong extent. This suggests that, in order to be successful against the 1.5C detector, the attack has to introduce a larger distortion into the image.

The attack success rate against the four SVMs is reported in Table \ref{tab:1-5C_Attack}, for all the detection tasks.
\begin{table}
	\caption{Percentage of misclassified attacked images. The attack is carried out against $1C_{H_0}^{cmb}$.}
	\label{tab:1-5C_Attack}   
	\begin{tabular}{lllll}
		\hline\noalign{\smallskip}
		& {\bf $2C_{H_{0/1}}$} & $1C_{H_0}$ & $1C_{H_1}$ & {\bf $1C_{H_0}^{cmb}$}  \\
		\noalign{\smallskip}\hline\noalign{\smallskip}
		Resizing & {\bf 100\%} & 1\% & 9\% & {\bf 100\%} \\
		Median Filter & {\bf 100\%} & 20\% & 21\% & {\bf 100\%}  \\
		CL-AHE & {\bf 100\%} & 23\% & 14\% & {\bf 100\%}\\
		\noalign{\smallskip}\hline
	\end{tabular}
\end{table}
Table \ref{tab:MSE} compares the attack against $2C_{H_{0/1}}$ and the 1.5C detector in terms of MSE. Specifically, the MSE averaged on the 100 attacked images in $\mathcal{S}_T^{t}$ is reported in the table for the two attacks.
We see that, in order to make the 1.5C classifier fail, the attacker must introduce a larger MSE with respect to the case in which the targeted classifier is $2C_{H_{0/1}}$: the average value of the MSE in the case of the attack against the 1.5C detector is more than twice that necessary for the case of resizing and contrast enhancement and almost double for the case of median filtering.
The average percentage of pixels modified by the two attacks are reported in Table \ref{tab:averagePixelNo}.
The table confirms that, in order to be successful against the 1.5C classifier, the attacker has to modify a larger number of pixels.
\begin{table}
	\caption{Average MSE.}
	\label{tab:MSE}  
	\begin{tabular}{llll}
		\hline\noalign{\smallskip}
		& Resize & Median Filter & CL-AHE  \\
		\noalign{\smallskip}\hline\noalign{\smallskip}
		Attack against $2C_{H_{0/1}}$ & 0.10 & 0.22 & 0.27  \\
		Attack against $1C_{H_0}^{cmb}$ & 0.17 & 0.60 & 0.43 \\
		\noalign{\smallskip}\hline
	\end{tabular}
\end{table}
\begin{table}
	\caption{Average percentage of pixels modified by the attack.}
	\label{tab:averagePixelNo}  
	\begin{tabular}{llll}
		\hline\noalign{\smallskip}
		& Resize & Median Filter & CL-AHE  \\
		\noalign{\smallskip}\hline\noalign{\smallskip}
		Attack against $2C_{H_{0/1}}$ & 9.5\% & 15.1\%  & 12.3\%  \\
		Attack against $1C_{H_0}^{cmb}$ & 13.4\% & 25.1\% & 15.1\% \\
		\noalign{\smallskip}\hline
	\end{tabular}
\end{table}
\subsubsection{Comparison with 2C classifier based on Convolutional Neural Networks (CNNs)}

In order to assess the performance of the proposed system with respect to state of the art CNN-based image manipulation detection, in this section, we consider the case of CNN-based 2C classification in the presence of a targeted attack.

For these experiments, we considered the CNN architecture in \cite{bayar2016deep}, used in the literature for several manipulation detection tasks.
	We set the input patch size to 128$\times$128. The datasets for training and testing the models were obtained by splitting into blocks the images in $S_{Tr}$ and $S_{V}$ for training (and validation) and $S_T$ for testing.
	The same parameters setting used in \cite{bayar2016deep} has been adopted for training the models (optimization solver, learning rate, batch size, etc \dots). All the models were trained on 20 epochs. The average test accuracy of the trained CNN models are 97.8\%, 81\% and  86.9\%  for resizing, median filtering and CL-AHE respectively.\footnote{The detection median filtering with a small ($3\times 3$) window, as well as the detection of CL-AHE are not easy tasks. Deeper networks could give better performance in this cases (see for instance \cite{barni2018cnn}); however, this goes in general at the price of lower robustness against attacks, as  deeper models are known to be more vulnerable to attacks than shallow ones \cite{szegedy2013intriguing}.}
%
Given a test  image,  the decision is made by dividing the image into patches (non-overlapping patches are considered for simplicity), testing each patch with the trained model, and then fusing the CNN outputs. For simplicity, the normalized sum of the decision scores ('0' for original, '1' for manipulated) is considered
	as the final score for the entire image. Then, for a given an image, the decision is made  by thresholding the accumulated score. The performance of the classification are measured again by relying on the ROC curve obtained by varying the decision threshold:  in particular we got AUC=1  for resizing, AUC =0.98 for median filtering and  AUC =0.93 for CL-AHE.

To attack the CNN classifiers, we considered the well known Jacobian-based Saliency Map Attack (JSMA) method  \cite{papernot2016limitations}, due to its good effectiveness even in the presence of integer rounding.
	To keep the distortion low, the attack is applied with the following setting: the relative amount of pixel modification ($\theta$) is set to 0.005; the maximum number of times the same pixel can be modified is set to 3; finally, the maximum number of iterations for the attack is set to 8000.
	We verified that similar performance are obtained by using the pixel-domain attack in \cite{TondiAttack} which extends the attack in \cite{Chen2017GraAttack} developed for the SVMs, to the case of CNNs.
	As before, the attack is applied to 100 of images from the test set $S_T^{t}$, belonging to the $H_1$ class. In order to be successful, the attack should be able to fool the CNN model, and then revert the decision, for at least half of the patches of the image.
	To minimize the overall distortion, the 'most favorable' patches are considered by the attack, that is, those patches that can be attacked by introducing  the minimum (MSE) distortion.

The final average MSE of the attack, averaged on all the 100 images, was:  0.055 for resizing, 0.094 for median filtering and 0.347 for CL-AHE.
	We observe that these MSE values are lower than those obtained with the 2C SVM  for the case of resize and median filtering detection, and always  lower  (significantly lower) than those for the 1.5C case (see Table \ref{tab:MSE}). This is not surprising, since it is known that CNNs are vulnerable to adversarial attacks and can be attacked by introducing  very small perturbations. Assessing the security gain that can be obtained by using CNNs to build a 1.5C classifier is an interesting piece of work, and will be considered as a future research.

\vspace{1.5cm}

\section{Conclusions}
\label{sec.con}

In this paper, we have proposed to use a multiple classifier architecture, referred to as 1.5C classifier, to  mitigate the damage made by an attacker with perfect knowledge acting against an image manipulation detector. In such a situation, the only possible defence for the analyst is to use a detector which is intrinsically more difficult to attack. This is the case of 1C classifiers, which, however, have the drawback of achieving inferior performance with respect to more conventional 2C classifiers. By properly combining one 2C classifier and three 1C classifiers, the adopted 1.5C classifier couples the advantages of 2C and 1C solutions, achieving a superior security while retaining the good performance of 2C classification in the absence of attacks. We implemented a particular instantiation of the proposed architecture by relying on four SVMs, and we trained it so to detect three kinds of image manipulations, namely median filtering, resizing and adaptive histogram equalization. The experimental analysis we carried out confirms that the 1.5C architecture is harder to attack than a 2C classifier with similar performance.

As possible directions for future works, we indicate the development of 1.5C solutions based on building blocks other than SVMs, for instance Convolutional Neural Networks (CNN) detectors or a mixture of CNNs and SVMs. The benefits achievable by training the 1.5C classifier with post-processed images so to improve its robustness against post-processing operators like noise addition, dithering or JPEG compression are also worth being investigated. Eventually, it would be interesting to train an adversary-aware 1.5C classifier, by including in the training phase some examples of attacked images.

\section*{Acknowledgments}
This work was supported partially by Defense Advanced Research Projects Agency (DARPA) and Air Force Research Laboratory (AFRL) under the research grant number FA8750-16-2-0173. The United States Government is certified to reproduce and distribute reprints for Governmental objectives notwithstanding any copyright notation thereon. The views and conclusions consist of herein are those of the authors and should not be explained as necessarily representing the official policies or authorization, either expressed or implied, DARPA and AFRL or U.S. Government.

\bibliographystyle{plain}
\bibliography{Ref}

\newpage

\begin{wrapfigure}{l}{43mm}
	\includegraphics[width=0.35\columnwidth,clip,keepaspectratio]{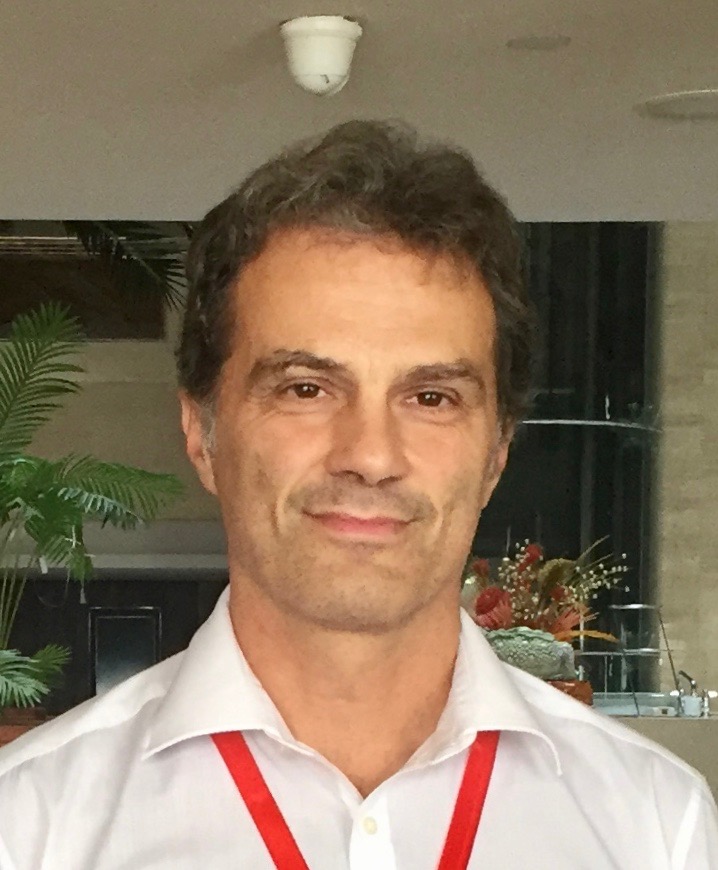}
\end{wrapfigure}\par
\noindent
\textbf{Mauro Barni} graduated in electronic engineering from the University of Florence in 1991. He received the Ph.D. degree in informatics and telecommunications in 1995. During the last two decades, he has been studying the application of image processing techniques to copyright protection and authentication of multimedia, and the possibility of processing signals that have been previously encrypted without decrypting them. Lately, he has been working on theoretical and practical aspects of adversarial signal processing.
He has authored or co-authored about 300 papers published in international journals and conference proceedings and holds five patents in digital watermarking and image authentication. He has co-authored the book \textit{Watermarking Systems Engineering: Enabling Digital Assets Security and other Applications} (Dekker, Inc., 2004). He participated in several National and European research projects on diverse topics, including computer vision, multimedia signal processing, remote sensing, digital watermarking, and IPR protection.
He is a member of EURASIP. He was a recipient of the Individual Technical Achievement Award of EURASIP for 2016. He has been the Chairman of the IEEE Information Forensic and Security Technical Committee from 2010 to 2011. He was the Technical Program Chair of ICASSP 2014. He was appointed DL of the IEEE SPS for the years 2013–2014. He was Editorin-Chief of the IEEE TRANSACTIONS ON INFORMATION FORENSICS AND SECURITY from 2015 to 2017. He was the Funding Editor of the \textit{EURASIP Journal on Information Security}. He has been serving as an Associate Editor of many journals, including several IEEE Transactions.\par

\newpage

\begin{wrapfigure}{l}{43mm}
	\includegraphics[width=0.36\columnwidth,clip,keepaspectratio]{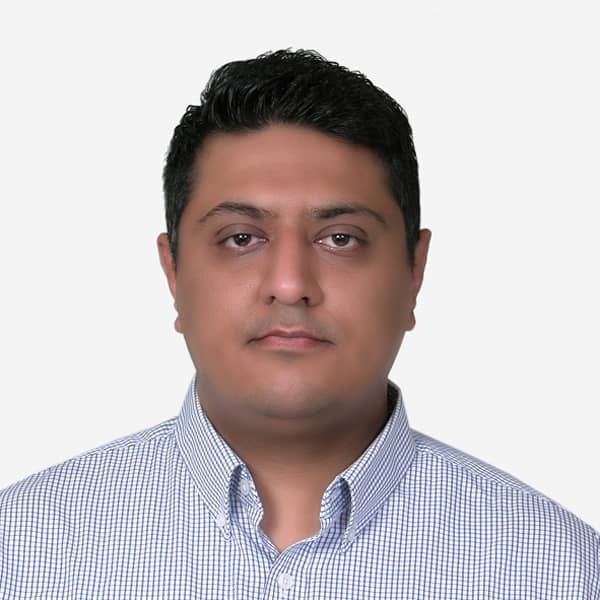}
\end{wrapfigure}\par
\noindent
\textbf{Ehsan Nowroozi} was born in Iran, Shiraz city in 10th of June 1986 and he is a  Ph.D. student at the University of Siena, Italy from 2016-2019.  His doctoral research investigates in Multimedia Forensics, with particular reference to: "Development of machine learning techniques for image and video forensics in an adversarial setting." He holds a master's degree in Computer Engineering - Computer Architecture from the Shahid Beheshti University (SBU), Tehran, Iran, that investigated "Double JPEG compression detection using statistical analysis." He graduated as the first rank during his M.Sc with GPA 17.95 out of 20.  In October 2016, he is selected as a Ph.D. student with the scholarship at the University of Siena, Italy in information engineering and mathematical sciences, working in the Visual Information Processing and Protection (VIPP) Lab under the supervision of Professor Mauro Barni. He wins one research scholarships, which supported by DARPA (Defense Advanced Reseach Projects Agency) and US Airforce Laboratory, USA. He has published various papers and a couple of books (Persian language) in the field of multimedia forensics. Also, he has been working on theoretical and practical aspects of adversarial multimedia forensics and adversarial machine learning with particular reference to the application of image processing techniques to authentication of multimedia (multimedia forensics). His professional service and activity as a reviewer of the journal of information security and applications - Elsevier from 24th of September, 2017 up to now and Elsevier journal Digital Investigation from November 2018 up to now. Also, he is a member of the IEEE Young Professionals and IEEE Signal Processing Society. \par

\newpage

\begin{wrapfigure}{l}{43mm}
	\includegraphics[width=0.35\columnwidth,clip,keepaspectratio]{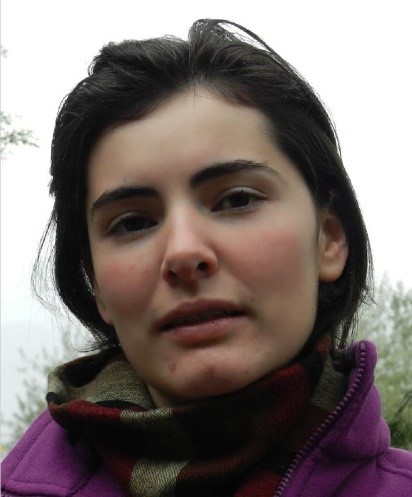}
\end{wrapfigure}\par
\noindent
\textbf{Benedetta Tondi} received the bachelor’s and master’s degree (cum laude) in electronics and communications engineering from the University of
Siena, Siena, Italy, in 2012, and the Ph.D. degree in 2016.
From 2014 to 2015, she was a Visiting Student with the Signal Processing in Communications Group, University of Vigo. She is currently a Research Associate with the Department of Information Engineering and Mathematics, University of Siena. She is an Assistant for the Course of Information Theory and Coding and Multimedia Security. Her research interest focuses on the application of information theory and game theory concepts to forensics and counter-forensics analysis and more in general on the adversarial signal processing, and on the use of deep learning techniques for multimedia forensics and security.
She is a member of the IEEE Young Professionals and IEEE Signal Processing Society and a member of the National Inter-University Consortium for Telecommunications (CNIT). Since 2019, she has been a member of the IEEE Information Forensics and Security Technical Committee. She is currently involved in the American DARPA MediFor Project and Media
Forensics Integrity Analysis Programme from 2016 to 2020. She was a recipient of the Best Student Paper Award at the IEEE International Workshop on Information Forensics and Security (WIFS) 2014, and the Best Paper Award at the IEEE International WIFS 2015 and at The Ninth International Conferences on Advances in Multimedia 2017. She was also a recipient of the 2017 GTTI Ph.D. Award for the Best Ph.D. Theses defended at an Italian University in the areas of communications technologies (signal processing, digital communications, and networking). Since 2018, she has been serving as an Associate Editor for the EURASIP Journal of Information Security. She was a Co-Organizer of a Special Session on Adversarial Multimedia Forensics at
EUSIPCO 2018. She has been a designated reviewer on the technical program committee of several IEEE Workshops an International Conferences. She is elected member of the Information Forensics and Security (IFS) Technical Committee (January 2019 - December 2021). \par
	
\end{document}